\documentclass[onecolumn,showpacs,preprintnumbers,amsmath,amssymb,preprint]{revtex4-1}
\usepackage{color}
\usepackage{graphicx}% Include figure files
\usepackage{dcolumn}% Align table columns on decimal point
\usepackage{bm}% bold math
\usepackage{ulem}% uline
\begin{document}

\preprint{}

\title{
Nucleon charges with dynamical overlap fermions
}% Force line breaks with \\

\newcommand{\ipno}{
IPNO, Universit\'{e} Paris-Sud, CNRS/IN2P3, F-91406, Orsay, France
}
\newcommand{\Rikenithes}{
iTHES Research Group, RIKEN, Wako, Saitama 351-0198, Japan
}
\newcommand{\FEFU}{
Complex Simulation Group, School of Biomedicine, Far Eastern Federal University, Vladivostok, 690950 Russia
}
\newcommand{\KEK}{
  KEK Theory Center,
  High Energy Accelerator Research Organization (KEK),
  Tsukuba 305-0801, Japan
}
\newcommand{\Rikenbnl}{
RIKEN BNL Research Center, Brookhaven National Laboratory, Upton New York 11973, USA
}
\newcommand{\NaraWU}{
Department of Physics, Nara Women's University, Nara 630-8506, Japan
}
\newcommand{\GUAS}{
  School of High Energy Accelerator Science,
  The Graduate University for Advanced Studies (Sokendai),
  Tsukuba 305-0801, Japan
}

\author{N.~Yamanaka}
  \email{yamanaka@ipno.in2p3.fr}
  \affiliation{\ipno}
  \affiliation{\Rikenithes}
  \affiliation{\FEFU}

\author{S.~Hashimoto}
\affiliation{\KEK}
\affiliation{\GUAS}

\author{T.~Kaneko}
\affiliation{\KEK}
\affiliation{\GUAS}

\author{H.~Ohki}
\affiliation{\NaraWU}
\affiliation{\Rikenbnl}

\collaboration{JLQCD Collaboration}
\noaffiliation

\date{\today}% It is always \today, today,
             %  but any date may be explicitly specified

\begin{abstract}

We calculate the scalar and tensor charges of the nucleon in 2+1-flavor lattice QCD, for which the systematics of the renormalization of the disconnected diagram is well controlled.
Numerical simulations are performed at a single lattice spacing $a=0.11$~fm.
We simulate four pion masses, which cover a range of $m_\pi \sim 290$\,--\,540~MeV, and a single strange quark mass close to its physical value.
The statistical accuracy is improved by employing the so-called low-mode averaging technique and the truncated solver method.
We study up, down, and strange quark contributions to the nucleon charges by calculating disconnected diagrams using the all-to-all quark propagator.
Chiral symmetry is exactly preserved by using the overlap quark action to avoid operator mixing among different flavors, which complicates the renormalization of scalar and tensor matrix elements and leads to possibly large contamination to the small strange quark contributions.
We also study the nucleon axial charge with a contribution from the disconnected diagram.
Our results are in reasonable agreement with experiments and previous lattice studies.

\end{abstract}

\pacs{11.15.Ha, 12.38.Gc, 14.20.Dh, 13.40.Em}% PACS, the Physics and Astronomy
                             % Classification Scheme.
%11.15.Ha --- Lattice gauge theory
%12.38.Gc --- Lattice QCD calculations
%14.20.Dh --- Protons and neutrons
%13.40.Em --- Electric and magnetic moments of hadrons

%\keywords{Suggested keywords}%Use showkeys class option if keyword
                              %display desired
\maketitle

\section{\label{sec:intro}Introduction}

The nucleon charges are very important input parameters in the study of new physics beyond the standard model, and accurate values are required in phenomenological analyses.
As a representative case, the nucleon scalar charge is important in the direct search for dark matters~\cite{darkmatter1,darkmatter2,darkmatter3,darkmatter4}.
The nucleon tensor charge relates the quark electric dipole moment to that of the nucleon, which is an important observable in the search for new sources of CP violation \cite{engel,atomicedmreview}.
The nucleon scalar and tensor charges are however difficult to directly measure in experiments, and no accurate experimental values are currently known.
They are thus important subjects to be studied in lattice QCD, since it is the only known method to calculate hadronic quantities with controlled uncertainties.

The nucleon charges have widely been studied in the literature.
The evaluation of the nucleon scalar charge in lattice QCD first began in the context of the investigation of the nucleon sigma term $\sigma_{\pi N } \equiv \sum_{q=u,d} \frac{m_q}{2m_N} \langle N| \bar q q | N \rangle$.
It is still a matter of debate due to the discrepancy between results of recent lattice QCD calculations at the physical pion mass, yielding values between 30 to 40 MeV \cite{qcdsf-ukqcdsigmaterm,chiqcdsigmaterm,etmsigmaterm,bmwsigmaterm,rqcdsigmaterm}, and phenomenological ones, giving almost 60 MeV \cite{alarcon,Hoferichter,Hoferichter2,yao,deelvira}.
The nucleon scalar charge also contains the isovector one as well as the strange content of the nucleon, which are now showing importance in the analysis of new physics beyond standard model.

The nucleon tensor charge is the leading twist contribution to the transversity distribution, one of the three parton distribution functions of the polarized nucleon \cite{barone}.
Currently, lattice QCD is the only way to accurately determine it.
Recent lattice calculations at the physical pion mass are giving tensor charges with a precision of 10\% \cite{rqcdisovector,pndmetensor1,pndmetensor2,etm2017}, while experimental and theoretical efforts to measure them beyond this accuracy are ongoing \cite{yez,accardi}.

We also note that the strange quark contribution to nucleon scalar and tensor charges is of particular interest.
This is because new physics of TeV scale or beyond, which contribute to low energy observables through those charges, are generated by interactions proportional to the strange quark mass, and consequently their effect is enhanced compared to that of light quarks.

The nucleon scalar and tensor charges require the flip of chirality of quarks.
In lattice QCD, those kinds of quantities generally suffer from an important systematics due to the renormalization of the disconnected diagram in the use of fermion action which explicitly breaks chiral symmetry.
To control this systematics, formulations which conserve chiral symmetry such as the domain-wall fermion or the overlap fermion are advantageous.
In Refs.~\cite{ohki,takeda,ohki2}, we exactly preserve chiral symmetry by using the overlap quark action~\cite{neuberger1,neuberger2}, and obtain $\sigma_s$ significantly smaller than previous estimates with the Wilson-type fermions~\cite{sigmatermandstrangelattice}.
This demonstrates the importance of controlling the systematics due to the explicit violation of chiral symmetry. 

In this paper, we present a comprehensive calculation of the nucleon scalar and tensor charges in $N_f\!=\!2+1$ QCD~\cite{hashimotochiu}.
Exact chiral symmetry preserved by the overlap action suppresses the operator mixing among different flavors~\cite{takeda}.
This simplifies the renormalization of the scalar and tensor charges, and allows us to avoid potentially large contamination to the small strange quark contributions from the light quark ones.
We exploited this advantage in the previous calculations of $\sigma_{\pi N}$ for $N_f\!=\!2$ through the Feynman-Hellmann theorem~\cite{ohki} and $\sigma_s$ for $N_f\!=\!2$ and 2+1 from nucleon three-point functions~\cite{takeda,ohki2}.
In this study, we extend these studies to separately calculate the up, down and strange quark contributions to the scalar and tensor charges.

We also calculate the nucleon axial charge which can be obtained in the same framework, since there are also good physical motivations.
The isovector axial charge ($g_A$) is a good benchmark with known experimental data \cite{ucna}, and it is also an important input of the chiral perturbation theory (ChPT).
The singlet axial charge is well known for posing the proton spin puzzle, where experimental data is showing a small contribution from quarks to the nucleon spin \cite{compass2}.
We calculate the above nucleon axial charges as well as the strange quark contribution, which may have important consequence in the theoretical research of supernova explosion \cite{melson,hobbs}.

To this end, relevant disconnected three-point functions of nucleon are calculated by using the all-to-all quark propagator~\cite{A2A:SESAM,A2A:TrinLat}.
To improve the statistical accuracy, we employ the low-mode averaging (LMA) technique~\cite{lma:ds,lma:ghlww} and the truncated solver method (TSM)~\cite{tsm}.
Preliminary results of this study were reported in Ref.~\cite{lat15:yamanaka}.

The structure of this paper is as follows.
In Section~\ref{sec:renormalization}, we discuss and show the importance of the chiral symmetry in the renormalization of the nucleon scalar and tensor charges.
Then in Section~\ref{sec:method}, we introduce details of our gauge ensembles and how to calculate the nucleon charges on them.
Our results for the axial, scalar and tensor charges are presented in Sections~\ref{sec:axial_analysis}, \ref{sec:scalar_analysis}, and \ref{sec:tensor_analysis}, respectively.
Our conclusions are summarized in Section~\ref{sec:conclusion}.

\section{Renormalization and chiral symmetry}
\label{sec:renormalization}

In this Section, we discuss the renormalization of the strange quark contribution to the nucleon scalar and tensor charges, since it is of crucial importance to consider the chiral symmetry in order to calculate them without large systematics.
We discussed in Ref.~\cite{takeda} that the disconnected contribution to renormalization of the nucleon scalar charge vanishes in a mass independent scheme provided that both the scheme and lattice simulation respect chiral symmetry.
The same argument also applies to the nucleon tensor charge.

We give a brief explanation of the renormalization of nucleon charges following Ref.~\cite{takeda}.
We define the renormalized quark bilinears of Dirac matrix $\Gamma \ (= \hat{1}, \gamma^\mu \gamma_5 , \sigma^{\mu \nu} )$ in the $SU(3)$ triplet basis $\psi^T \equiv (u,d,s)$ as 
\begin{eqnarray}
(\bar \psi \Gamma \psi )_{\rm phys}
&=&
Z_{\Gamma 0} (\bar \psi \Gamma \psi ) 
,
\\
(\bar \psi \Gamma \lambda_3 \psi )_{\rm phys}
&=&
Z_{\Gamma 3} (\bar \psi \Gamma \lambda_3 \psi ) 
,
\\
(\bar \psi \Gamma \lambda_8 \psi )_{\rm phys}
&=&
Z_{\Gamma 8} (\bar \psi \Gamma \lambda_8 \psi ) 
,
\end{eqnarray}
where $\lambda_3$ and $\lambda_8$ are the Gell-Mann matrices.
The singlet ($Z_{\Gamma 0}$) and nonsinglet ($Z_{\Gamma 3}, Z_{\Gamma 8}$) renormalization factors are not identical in the general case.
For the case of the scalar charge, the singlet operator can also mix with the identity operator.
By focusing on the renormalized strange quark bilinear, its general expression is then given by
\begin{equation}
(\bar s \Gamma s )_{\rm phys}
=
\frac{1}{3}
\Biggl[
(Z_{\Gamma 0} + 2 Z_{\Gamma 8} ) (\bar s \Gamma s )
+
(Z_{\Gamma 0} - Z_{\Gamma 8} ) (\bar u \Gamma u +\bar d \Gamma d)
+ \frac{b_0}{a^3}
+\cdots
\Biggr]
,
\label{eq:ssbareq}
\end{equation}
where $b_0$ is a constant which is only nonzero for the scalar charge.

\begin{figure}[tb]
\begin{center}
\includegraphics[width=12cm]{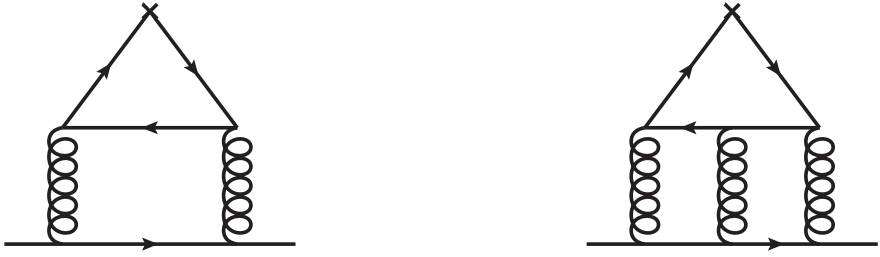}
\caption{\label{fig:renormalization_disconnected} 
The disconnected diagram contribution to the renormalization of singlet nucleon scalar, axial (left) and tensor (right) charges.
If the chiral symmetry is preserved in the regularization, this quark-loop contribution to the renormalization of the scalar and tensor charges vanishes since the quark-quark-gluon vertex does not change the chirality.
The quark loop of the tensor charge (right) requires at least three gluons in order to satisfy Furry's theorem.
The crosses on the top of each diagram denote the quark charge operator ($\Gamma = \hat{1}, \gamma^\mu \gamma_5 , \sigma^{\mu \nu} $).
}
\end{center}
\end{figure}

From Eq. (\ref{eq:ssbareq}), we see that $\bar s \Gamma s$ mixes with $\bar u \Gamma u +\bar d \Gamma d$ if $Z_{\Gamma 0} - Z_{\Gamma 8} \ne 0$.
The term with $Z_{\Gamma 0} - Z_{\Gamma 8}$ in Eq. (\ref{eq:ssbareq}) is actually given by the disconnected diagram, since it is the difference between the singlet and the nonsinglet operators.
In the mass independent renormalization scheme (like the $\overline{\rm MS}$ scheme), the disconnected diagram contribution to the renormalization of the nucleon scalar and tensor charges vanishes, since these operators have to change the chirality in the quark loop (see Fig. \ref{fig:renormalization_disconnected}).
Consequently, we have $Z_{\Gamma 0} = Z_{\Gamma 8} (= Z_{\Gamma 3} \equiv Z_{\Gamma })$, so that $\bar s \Gamma s$ does not mix with $\bar u \Gamma u +\bar d \Gamma d$ and we only need to calculate one renormalization factor.
For the scalar charge, the cancellation of the quark-loop also prohibits it to contribute to the vacuum and therefore the divergent term $\frac{b_0}{a^3}$ of Eq. (\ref{eq:ssbareq}) also cancels.
This cancellation is guaranteed in the overlap fermion formulation, where the Ginsparg-Wilson relation holds at finite lattice spacing \cite{luscher}.

In contrast, the chiral symmetry is explicitly broken for the conventional Wilson fermion formulation.
In that case, we have to separately calculate the nonperturbative renormalization factors $Z_{\Gamma 0}$ and $Z_{\Gamma 8}$, since they are not equal.
The finite $Z_{\Gamma 0} - Z_{\Gamma 8}$ then induces a mixing between $\bar s \Gamma s$ and $\bar u \Gamma u +\bar d \Gamma d$~\cite{Michael:2001bv,ohki,takeda}.
Because $\bar u \Gamma u +\bar d \Gamma d$ contains a large connected contribution, $\bar s \Gamma s$ may receive a sizable contamination from $\bar u \Gamma u +\bar d \Gamma d$ even for small $Z_{\Gamma 0} - Z_{\Gamma 8}$, and this makes it difficult to extract the true signal of $\bar s \Gamma s$.
For the scalar charge, moreover, the divergent term $\frac{b_0}{a^3}$ has to be subtracted as part of the vacuum expectation value of $\bar{s}s$.
These difficulties are carefully considered in modern calculations:
{\it e.g.} Refs.~\cite{qcdsf-ukqcdsigmaterm,rqcdsigmaterm,etmsigmaterm,etm2017}
for the scalar charges,
and see Ref.~\cite{Young:2009ps} for a comparison of earlier lattice results.
The overlap action, however, provides a straightforward calculation
of the isoscalar and strange quark contributions
to the scalar and tensor charges. 

We also comment on the renormalization of the nucleon axial charge.
For this case, the contribution of the disconnected diagram to the renormalization factor does not vanish.
We then have $Z_{A 0} - Z_{A 8} \ne 0$, and $\bar s \gamma^\mu \gamma_5 s$ mixes with $\bar u \gamma^\mu \gamma_5 u + \bar d \gamma^\mu \gamma_5 d$ which contains a relatively large connected contribution.
Since there is no result of calculations of $Z_{A 0} $ on lattice, we cannot evaluate its strange quark contribution without large uncertainty.
For the singlet axial charge, however, we can evaluate it, since it is dominated by the contribution from the connected diagram.
The disconnected contribution to $Z_{A0}$ may be non-negligible, but is ignored in this study. 
We note that such contribution turns out to be not large ($\lesssim 5$~\%) in a perturbative analysis for the Wilson-type fermions \cite{qcdsfprotonspin,Skouroupathis,etmpertrenor}.
The nonperturbative estimate for the twisted mass fermions gives $Z_{A3}=0.7910(4)(5)$ and $Z_{A0}=0.7968(25)(91)$ \cite{etmprotonspin,etmaxial2017}, a discrepancy which is consistent with zero within the error bar.
Another potential contribution to the renormalization of the axial charge is the nonperturbative effect due to the topological number of the gauge configuration.
This latter will be evaluated later separately.

\section{Simulation method}
\label{sec:method}

\subsection{Simulation setup and gauge ensembles}

We simulate $N_f=2+1$ flavor QCD using 
the overlap quark action~\cite{neuberger1,neuberger2}.
Its Dirac operator is given by 
\begin{equation}
D (m) =
\Bigl( m_0 + \frac{m}{2} \Bigr)
+ \Bigl( m_0 - \frac{m}{2} \Bigr) \gamma_5 {\rm sgn} [H_W]
,
\label{eq:overlap_dirac_operator}
\end{equation}
where $m$ represents the quark mass,
and $H_W = \gamma_5 D_W (- m_0 )$ is the Hermitian Wilson-Dirac operator.
A large negative mass $-m_0 = -1.6$ is chosen so that 
$D(m)$ has good locality.

For the gauge fields, we employ the Iwasaki gauge action~\cite{Iwasaki} 
including a term 
$\delta S_g\!=\!-\mbox{ln}\left[ \Delta \right]$
with~\cite{exW}
\begin{equation}
\Delta 
=
{\rm det} \Biggl[
\frac{H_W (-m_0)^2}{H_W (-m_0)^2 + \mu^2}
\Biggr]
\hspace{5mm} (\mu\!=\!0.2)
.
\end{equation}
This modification does not change the  continuum limit of the theory, but remarkably accelerates our simulation by suppressing (near-)zero modes of $H_W$.
While the {\it global} topological charge $Q$ is fixed with the commonly used Hybrid Monte Carlo algorithm, its effects are suppressed by the inverse space-time volume~\cite{fixed_topology}.
Indeed, the $Q$ dependence turned out to be insignificant in our data of the pion form factors~\cite{JLQCD:Nf2:PFF} with a better accuracy than that for the nucleon observables.  
In this study, we mainly simulate the trivial topological sector with $Q\!=\!0$.
We also carry out an auxiliary simulation at $Q\!=\!1$ to check the effects of fixed $Q$ to the singlet axial charge $\Delta \Sigma$, which has the same quantum number as $Q$.

The bare gauge coupling is set to $\beta \equiv 6/g^2 =2.3$, where the lattice spacing fixed from the $\Omega$ baryon mass is $a=0.112 (1)$ fm.
We work in the isospin symmetric limit, and take four values $m_{ud} = 0.015 , 0.025 , 0.035$ and 0.050 for the mass of degenerate up and down quarks.
This choice covers the pion masses $m_\pi \simeq 290$\,--\,540~MeV, and $m_{ud} =0.0029$ corresponds to the physical pion mass $m_{\pi,{\rm phys}}$.
The strange quark mass is fixed to $m_s = 0.080$, which is very close to the physical value $m_{s, {\rm phys}} = 0.081$ fixed from the kaon mass $m_K$.
The $m_s$ dependence of the nucleon observables is negligibly small compared to our accuracy.

Depending on $m_{ud}$, we choose a lattice volume, $N_s^3 \times N_t = 16^3 \times 48$ or $24^3 \times 48$, to fulfill the condition $m_\pi L \geq 4$ for the control of finite volume effects due to pions wrapping around the lattice.
The statistics are 50 gauge configurations at each $m_{ud}$. 
Our simulation parameters are summarized in Table~\ref{table:simulation_parameter}.

\begin{table}[htb]
\begin{center}
\begin{tabular}{l|ccc}
\hline \hline
$m_{ud}$ & $m_\pi$~[MeV] & Lattice size\\
\hline
0.050 & 540(4) & $16^3 \times 48$ \\
0.035 & 453(4) & $16^3 \times 48$ \\
0.025 & 379(2) & $24^3 \times 48$ \\
0.015 & 293(2) & $24^3 \times 48$ \\
\hline
\end{tabular}
\end{center}
\caption{
Parameters of our simulations.}
\label{table:simulation_parameter}
\end{table}

\subsection{Calculation of nucleon charges}

The nucleon scalar, axial and tensor charges are defined by 
\begin{eqnarray}
S_q 
&\equiv&
\frac{1}{2 m_N} \langle N | \bar q q | N \rangle
,
\label{eq:scalar_charge}
\\
\Delta q 
&\equiv&
\frac{1}{2 m_N} \langle N (s_z = +1/2) | \bar q \gamma_3 \gamma_5 q | N (s_z = +1/2) \rangle
,
\label{eq:axial_charge}
\\
\delta q 
&\equiv&
\frac{1}{2 m_N} \langle N (s_z = +1/2) | \bar q i \sigma_{03} \gamma_5 q | N (s_z = +1/2) \rangle
,
\label{eq:tensor_charge}
\end{eqnarray}
respectively.
In this study,
we focus on the proton charges ($N=p$) and separately calculate up, down and strange quark contributions ($q=u,d,s$).
Note that the proton is polarized for the axial and tensor charges.

These charges can be extracted from the nucleon two- and three-point functions defined as
\begin{eqnarray}
C_{{\rm 2pt}} ( t_{\rm src} ,   {\bf y}_{\rm src},  \Delta t^\prime)
&=&
\frac{1}{N_s^6} \sum_{{\bf x}  }
{\rm tr}_s 
\Bigl[
\Gamma_+ 
\bigl<
N ({\bf x } , t_{\rm src}+\Delta t^\prime ) \bar N ( {\bf y}_{\rm src} , t_{\rm src} )
\bigr>
\Bigr]
,
\label{eq:2ptfunction}
\\
C_{{\rm 3pt}} ( t_{\rm src} ,   {\bf y}_{\rm src}, \Delta t,  \Delta t^\prime)
&=&
\frac{1}{N_s^6} \sum_{{\bf x} ,{\bf z} }
\Biggl\{
{\rm tr}_s 
\Bigl[
\Gamma_+ P_i 
\bigl<
N ({\bf x } , t_{\rm src}+\Delta t^\prime ) 
{\mathcal O}_{\Gamma_i} ({\bf z} , t_{\rm src}+\Delta t) \bar N ( {\bf y}_{\rm src} , t_{\rm src} )
\bigr>
\Bigr]
\nonumber\\
&& \hspace{3em}
-
\bigl<
{\rm tr}_s 
\bigr[
{\mathcal O}_{\Gamma_i} ({\bf z} , t_{\rm src}+\Delta t)
D^{-1}[({\bf z}, t_{\rm src}+\Delta t) , ({\bf z}, t_{\rm src}+\Delta t)]
\bigl]
\nonumber\\
&& \hspace{8em} \times
{\rm tr}_s 
\bigl[
\Gamma_+ P_i 
N ({\bf x } , t_{\rm src}+\Delta t^\prime) \bar N ( {\bf y}_{\rm src} , t_{\rm src} )
\bigr]
\bigr>
\Biggr\}
,
\label{eq:3ptfunction}
\end{eqnarray}
where the nucleon interpolating operator is given by $N = \epsilon_{abc} (u_a^T C \gamma_5 d_b) u_c$, and $(t_{\rm src},{\bf y}_{\rm src})$ is the location of the nucleon source operator $\bar N$.
We denote the temporal separation between the nucleon sink operator (charge operator ${\mathcal O}_{\Gamma_i}$) and $\bar N$ by $\Delta t'$ ($\Delta t$).
Their spatial coordinates, ${\bf x}$ and ${\bf z}$, are summed over the spatial volume to set the initial and final nucleon momenta to zero. 
$\Gamma_\pm = \frac{1}{2} (1\pm \gamma_0)$ is the projector to the nucleon propagating forward/backward in time.
For the axial and tensor charges, we polarize the nucleon with three possible directions ($x$, $y$, and $z$), namely we insert a projector $P = \frac{1}{2} (1+ \gamma_5 \gamma_i ) -\frac{1}{2} (1- \gamma_5 \gamma_i ) = \gamma_5 \gamma_i $, with $i=1,2,3$.
The quark axial and tensor charge operators are then set to ${\mathcal O}_{A_i}\!=\!\bar{q}\gamma_i \gamma_5 q$ and ${\mathcal O}_{T_{0i}}\! = i\bar{q}\sigma_{0i}\gamma_5 q$, respectively.
For the scalar charges, ${\mathcal O}_S\!=\!\bar{q}q$, and $P_i=1$ since they do not depend on the polarization of the nucleon.
To improve the statistical accuracy, the correlation functions with $\Gamma_+$ and $\Gamma_-$ are averaged by replacing $(\Delta t , \Delta t')$ by $(-\Delta t , -\Delta t')$ and by taking the charge conjugation of Dirac matrices.

\begin{figure}[ht]
\begin{center}
\includegraphics[width=12cm]{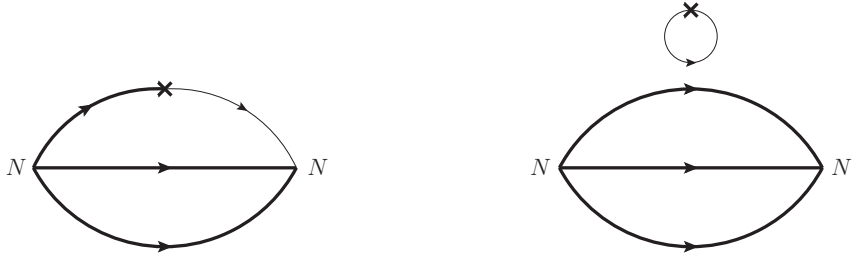}
\caption{\label{fig:connected_disconnected_diagrams} 
The connected (left) and disconnected (right) contributions to the nucleon charges.
The crosses denote the quark charge operator ($\Gamma = \hat{1} , \gamma^\mu \gamma_5, \sigma^{\mu \nu} $).
We use the point-to-all and all-to-all propagators for the quark propagators shown by thick and thin lines, respectively.
Note that, for the connected diagram, there are also diagrams with the charge operator inserted into one of the other two quark propagators.
}
\end{center}
\end{figure}

The three-point function $C_{\rm 3pt}$ is made of two contributions, namely the first and second terms in Eq.~(\ref{eq:3ptfunction}), which come from the connected and disconnected diagrams shown in Fig~\ref{fig:connected_disconnected_diagrams}.
A conventional way to calculate the quark propagator in these diagrams is to solve the linear equation
\begin{equation}
\sum_{x^\prime} D(x,x^\prime)\psi_{\rm pt}(x^\prime) = b(x),
\hspace{5mm}
b(x) = \delta_{{\bf x},{\bf y}_{\rm src}}\delta_{x_4,t_{\rm src}}.
\label{eqn:p2a_prop}
\end{equation}
Since $\psi_{\rm pt}(x)$ represents a quark propagator that flows from a given source point $({\bf y}_{\rm src},t_{\rm src})$ to any lattice site $x$, it is referred to as the point-to-all propagator in the following.
We use this type of quark propagator for the thick lines in Fig.~\ref{fig:connected_disconnected_diagrams} as well as to calculate the two-point function $C_{\rm 2pt}$.

For the momentum projection, $C_{\rm 3pt}$ has to be summed over ${\bf z}$ in Eq.~(\ref{eq:3ptfunction}), which is the source point of the quark propagator shown by thin lines in the same figure.
To this end, we need the all-to-all quark propagator, which flows from any to any lattice sites.
Since it is prohibitively time consuming to calculate $\psi_{\rm pt}$ for any source points, we calculate the all-to-all quark propagator by using deflation and stochastic methods~\cite{A2A:SESAM,A2A:TrinLat}.

Let us decompose the all-to-all propagator into the contribution of low-lying modes of the overlap-Dirac operator~(\ref{eq:overlap_dirac_operator}) and the remaining high-mode contribution.
The former is exactly calculated as 
\begin{equation}
(D^{-1} )_{\rm low}  (x,y)
=
\sum_{i=1}^{N_e}
\frac{1}{\lambda^{(i)}}
v^{(i)} (x) v^{(i) \dagger} (y),
\label{eq:lowmode}
\end{equation}
where $\lambda^{(i)}$ and $v^{(i)}$ denote the $i$-th lowest eigenvalue of $D$ and the corresponding eigenvector, respectively.
In this study, we use $N_e\!=\!160$ (240) low-lying modes on the $16^3 \times 48$ ($24^3 \times 48$) lattice.

The high-mode contribution is stochastically estimated by using the noise method~\cite{noise}.
We prepare a complex $Z_2$ noise vector $\eta (x)$ for each configuration, and split it into $N_d = 3\times 4 \times N_t / 2$ vectors $\eta^{(d)} (x)$ $(d\!=\!1,\cdots,N_d)$, which have non-zero elements only for a single combination of color and spinor indices on two consecutive time slices~\cite{takeda,ohki2}.
The high-mode contribution is then calculated as 
\begin{equation}
(D^{-1} )_{\rm high}  (x,y)
=
\sum_{d=1}^{N_d}
\psi^{(d)} (x) \eta^{(d) \dagger} (y) 
,
\label{eq:highmode}
\end{equation}
where $\psi^{(d)} (x)$ is the solution of $D \psi^{(d)} \!=\! (1-P_{\rm low}) \eta^{(d)}$ with $P_{\rm low}$ the projection operator into the eigenspace spanned by the low modes $\{v^{(i)}\}$.

The nucleon charges are extracted from the asymptotic behavior of $C_{\rm 3pt}$ and $C_{\rm 2pt}$ towards the limit of $\Delta t,\Delta t^\prime-\Delta t\!\to\!\infty$, where the ground state contribution becomes dominant.
The relevant nucleon matrix elements in the right-hand sides of Eqs.~(\ref{eq:scalar_charge})\,--\,(\ref{eq:tensor_charge}) are calculated from the following ratio
\begin{equation}
R_\Gamma (\Delta t, \Delta t^\prime)=
Z_\Gamma \frac{C_{\rm 3pt}( \Delta t, \Delta t^\prime)}{C_{\rm 2pt}( \Delta t^\prime)}
\xrightarrow[\Delta t,\Delta t^\prime - \Delta t \to \infty\\ ]{ } 
\frac{\langle N | Z_\Gamma {\mathcal O}_\Gamma | N \rangle}{2m_N},
\label{eq:nucleoncharge}
\end{equation}
where $Z_\Gamma$ represents the renormalization factor of the charge operator ${\mathcal O}_{\Gamma}$.
This study employs our estimate in Ref.~\cite{noaki} for the flavor nonsinglet operators in the $\overline{\rm MS}$ scheme at the scale $\mu\!=\!2$~GeV.
As discussed in Sec. \ref{sec:renormalization}, the same $Z_S$ and $Z_T$ can also be used for flavor singlet operators.
We also neglect the correction to $Z_A$ which is expected to be small.

To improve the statistical accuracy of $C_{\rm 3pt}$ and $C_{\rm 2pt}$, they are averaged over the source location $(t_{\rm src},{\bf y}_{\rm src})$ (see Sec.~\ref{sec:method:impr} for details).
We also suppress the excited state contamination to $C_{\rm 3pt}$ and $C_{\rm 2pt}$ by employing the Gaussian smearing to the quark fields in the nucleon interpolating operator $N$
\begin{equation}
q_{\rm smear} ({\bf x } , t) =
\sum_{\bf y }
\left\{ 
\left( 1+ \frac{\omega}{4N} H \right)^N
\right\}_{{\bf x } ,{\bf y }}
q_{\rm local} ({\bf y } , t)
,
\hspace{5mm}
H_{{\bf x},{\bf y}}
=
\sum_{i=1}^3 (\delta_{{\bf x},{\bf y}-\hat{\bf i}} + \delta_{{\bf x},{\bf y}+\hat{\bf i}}).
\label{eq:gaussian_smearing}
\end{equation}
Here we omit the gauge link, which enhances the statistical fluctuation of $C_{\rm 3pt}$ and $C_{\rm 2pt}$. This smearing is therefore gauge noninvariant, and we fix the gauge to the Coulomb gauge.
The parameters $\omega\!=\!20$ and $N\!=\!400$ are chosen in Ref.~\cite{takeda} by inspecting the excited state contamination to the effective mass of $C_{\rm 2pt}$.

\subsection{\label{sec:method:impr}Improvement of the statistical accuracy}

The all-to-all quark propagator is useful to calculate both connected and disconnected diagrams, and we have successfully applied it for the precision calculation of light meson matrix elements~\cite{JLQCD:Nf2:PFF,JLQCD:Nf3:EMFF,JLQCD:Nf3:K2pi}.
With our setup, however, it introduces relatively large statistical fluctuation to the nucleon correlators, which rapidly decay as $\propto \exp[ -M_N \Delta t^\prime ]$, since only a single noise sample is taken for each configuration.
In order to improve the statistical accuracy, the correlation functions are decomposed as 
\begin{equation}
C_{\rm 2pt}
=
C_{\rm 2pt,low} + C_{\rm 2pt,high},
\hspace{5mm}
C_{\rm 3pt}
=
C_{\rm 3pt,low} + C_{\rm 3pt,high},
\end{equation}
where $C_{\rm 2pt(3pt),low}$ represents the contribution in which the low-mode truncation (\ref{eq:lowmode}) is used for all quark propagators, and $C_{\rm 2pt(3pt),high}$ is the remaining contribution.
We suppress the statistical fluctuation of these contributions {\it \`{a} la} all-mode averaging technique~\cite{blumama}.

For $C_{\rm 3pt,low}$ and $C_{\rm 2pt,low}$, we employ the low-mode averaging (LMA) technique~\cite{lma:ds,lma:ghlww}.
Relying on the translational invariance, we replace these contributions by a more precise estimate, which is averaged over different source points $(t_{{\rm src},i},{\bf y}_{{\rm src},i})$ ($i=1,\ldots,N_{\rm LMA}$).
This can be expressed as 
\begin{eqnarray}
C_{\rm 2pt,low}\left(t_{\rm src}, {\bf y}_{\rm src}, \Delta t^\prime\right)
&\to & 
C_{\rm 2pt,low}^{\rm (LMA)}
\left( \Delta t^\prime\right)
=
\frac{1}{N_{\rm LMA}}
\sum_{i=1}^{N_{\rm LMA}}
C_{\rm 2pt,low} \left(t_{{\rm src},i}, {\bf y}_{{\rm src},i}, \Delta t^\prime \right)
,
\\
C_{\rm 3pt,low}\left(t_{\rm src}, {\bf y}_{\rm src}, \Delta t, \Delta t^\prime\right)
&\to & 
C_{\rm 3pt,low}^{\rm (LMA)}
\left(\Delta t, \Delta t^\prime\right)
=
\frac{1}{N_{\rm LMA}}
\sum_{i=1}^{N_{\rm LMA}}
C_{\rm 3pt,low} \left(t_{{\rm src},i}, {\bf y}_{{\rm src},i}, \Delta t, \Delta t^\prime \right)
.
\ \ \ \ \ 
\end{eqnarray}
The number of source points is chosen as a compromise between the statistical accuracy and computational cost.
It is $N_{\rm LMA}\!=\!48$ and 96 for the connected contribution to the axial and tensor charges at $m_{ud}\!\leq\!0.025$ and $\geq\!0.035$, respectively.
It is increased to 192 source points for the noisy scalar charge.
The disconnected contributions are much noisier, as we will see.
We take a rather large number $N_{\rm LMA}\!=\!768$ for these contributions and $C_{\rm 2pt}$, which is the nucleon piece of the disconnected diagram.
We also average the low-mode contribution of the nucleon in the disconnected diagram over three possible spatial directions to effectively increase the statistics.

We employ the TSM~\cite{tsm} to improve the statistical accuracy of
the high-mode contributions, which are replaced by a more precise estimate
\begin{eqnarray}
C_{\rm 2pt,high}^{\rm (TSM)}
\left( \Delta t^\prime \right)
& = &
C_{\rm 2pt,high}\left(1, {\bf 1}, \Delta t^\prime\right)
-\tilde{C}_{\rm 2pt,high}\left(1, {\bf 1}, \Delta t^\prime\right)
\nonumber \\
& & 
\hspace{10mm}
+ 
\frac{1}{{N_{\rm TSM}}}
\sum_{i=1}^{N_{\rm TSM}} 
\tilde{C}_{\rm 2pt,high} \left(t_{{\rm src},i}, {\bf y}_{{\rm src},i}, \Delta t^\prime \right)
,
\label{eq:2ptTSM}
\\
C_{\rm 3pt,high}^{\rm (TSM)}
\left(\Delta t, \Delta t^\prime \right)
& = &
C_{\rm 3pt,high}\left(1, {\bf 1}, \Delta t, \Delta t^\prime\right)
-\tilde{C}_{\rm 3pt,high}\left(1, {\bf 1}, \Delta t, \Delta t^\prime\right)
\nonumber \\
& & 
\hspace{10mm}
+ 
\frac{1}{{N_{\rm TSM}}}
\sum_{i=1}^{N_{\rm TSM}}
\tilde{C}_{\rm 3pt,high} \left(t_{{\rm src},i}, {\bf y}_{{\rm src},i}, \Delta t, \Delta t^\prime \right)
,
\label{eq:3ptTSM}
\end{eqnarray}
where $(1,{\bf 1})$ denotes the origin of the lattice.
The point-to-all quark propagators in $C_{\rm 2pt(3pt),high}$ 
in the right-hand sides
are calculated by solving Eq.~(\ref{eqn:p2a_prop})
with a strict stopping condition $|D \psi_{\rm pt} - b| \leq 10^{-7}$.
We use a more relaxed condition $|D \psi_{\rm pt} - b| \leq 10^{-2}$
for $\tilde{C}_{\rm 2pt(3pt),high}$ to average them 
over many source points $(t_{{\rm src},i},{\bf y}_{{\rm src},i})$
$(i\!=\!1,\ldots,N_{\rm TSM})$.
The number of source points is $N_{\rm TSM} = 48$ at $m_{ud}\!\leq\!0.025$,
and is increased to 96 for computationally inexpensive simulations 
at $m_{ud}\!\geq\!0.035$.

\begin{figure}[tbp]
\centering
\includegraphics[width=0.48\textwidth,clip]{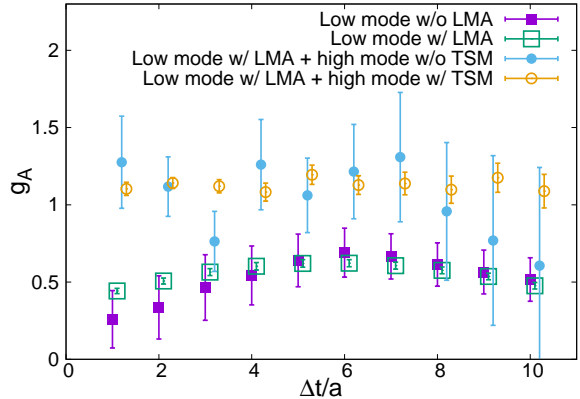}
\caption{\label{fig:lmaama}
Improvement of statistical accuracy of isovector axial charge $g_A$ by using LMA and TSM. 
We plot the effective value of $g_A$ at $m_{ud}=0.015$ as a function of $\Delta t$.
Open and filled squares show the low-mode contributions to $g_A$ with and without LMA, respectively, whereas open and filled circles include the high-mode contribution with and without TSM.
}
\end{figure}

The improvement of the statistical accuracy is demonstrated in Fig.~\ref{fig:lmaama}, where we plot the effective value of the isovector axial charge $g_A$ determined from the ratio $R_A(\Delta t,\Delta t^\prime=11)$ [Eq. (\ref{eq:nucleoncharge})] at $m_{ud}\!=\!0.015$.
The open and filled squares are the low mode contributions calculated by using $C_{\rm 3pt,low}$ and $C_{\rm 2pt,low}$ in $R_A$.
We observe a remarkable improvement of the statistical accuracy by a factor of 7. This is close to the ideal value $\sqrt{N_{\rm LMA}}\!=\!\sqrt{48}$ suggesting relatively small correlation among different source points with this choice of $N_{\rm LMA}$ on the $24^3\!\times\!48$ lattice.

\begin{figure}[tp]
\centering
\includegraphics[width=0.48\textwidth,clip]{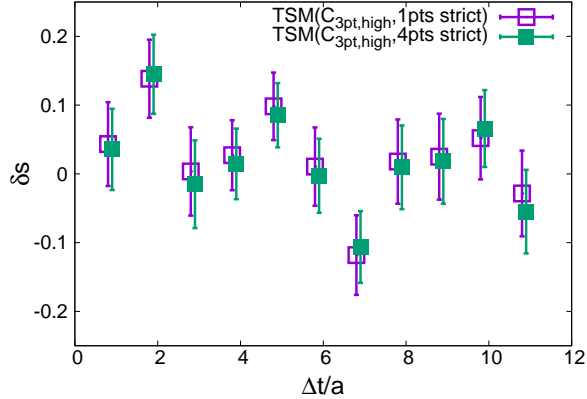}
\caption{\label{fig:TSMP2A4pts}
Strange quark contribution to the nucleon tensor charge, $\delta s$, 
calculated with TSM setups (\ref{eq:3ptTSM}) (open square) and
(\ref{eq:3ptTSM:4strict}) (solid square).
We plot the data at $m_{ud}\!=\!0.035$.
}
\end{figure}

\begin{figure}[bp]
\centering
\includegraphics[width=0.48\textwidth,clip]{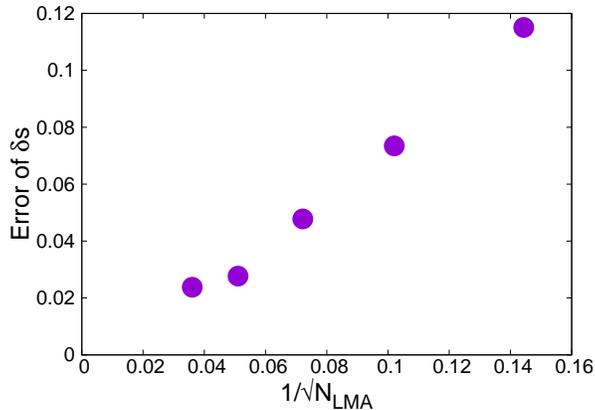}
\caption{\label{fig:discerror}
Statistical error of $\delta s$ as a function of $1/\sqrt{N_{\rm LMA}}$. 
Here we projected out the high modes from the quark propagator of the nucleon in $C_{\rm 3pt}$.
We plot data at $m_{ud}\!=\!0.015$.
}
\end{figure}

By applying the LMA, the statistical error of the full contribution is dominated by that of the high-mode contribution.
The same figure also shows that this error is largely reduced by using the TSM.
We typically observe an improvement by a factor of five, which is also close to $\sqrt{N_{\rm TSM}}\!=\!\sqrt{48}$,
while the TSM increases the computational cost only by a factor of 5
$\ll N_{\rm TSM}$.

In the TSM (\ref{eq:2ptTSM}) and (\ref{eq:3ptTSM}),
one may calculate $C_{\rm 2pt,high}$ and $C_{\rm 3pt,high}$
with precise quark propagators at multiple source points.
For the three-point function $C_{\rm 3pt,high}^{\rm (TSM)}$,
for instance, 
the right-hand side of Eq.~(\ref{eq:3ptTSM}) is replaced as 
\begin{eqnarray}
\frac{1}{4}\sum_{j=1}^{4}
C_{\rm 3pt,high}
\left(t_{{\rm src},j}^{\prime}, {\bf x}_{{\rm src},j}^{\prime}, \Delta t, \Delta t^\prime\right)
& - & 
\frac{1}{4}\sum_{j=1}^{4}
\tilde{C}_{\rm 3pt,high}\left(t_{{\rm src},j}^{\prime}, {\bf x}_{{\rm src},j}^{\prime}, \Delta t, \Delta t^\prime\right)
\nonumber \\
& 
+
& 
\frac{1}{{N_{\rm TSM}}}
\sum_{i=1}^{N_{\rm TSM}} 
\tilde{C}_{\rm 3pt,high} \left(t_{{\rm src},i}, {\bf y}_{{\rm src},i}, \Delta t, \Delta t^\prime \right),
\label{eq:3ptTSM:4strict}
\end{eqnarray}
where we calculate $C_{\rm 3pt,high}$ at four source points
$(t_{{\rm src},j}^{\prime}, {\bf x}_{{\rm src},j}^{\prime})$ ($j=1,...,4$).
Figure~\ref{fig:TSMP2A4pts} compares
the strange quark contribution to the tensor charge, $\delta s$,
with the nucleon propagator
calculated through the TSM (\ref{eq:3ptTSM}) and (\ref{eq:3ptTSM:4strict}).
The statistical errors are comparable to each other.
This suggests that our choice (\ref{eq:2ptTSM}) and (\ref{eq:3ptTSM})
is reasonably good,
while it has the least cost to calculate precise quark propagators.

\begin{figure}[bp]
\centering
\includegraphics[width=0.48\textwidth,clip]{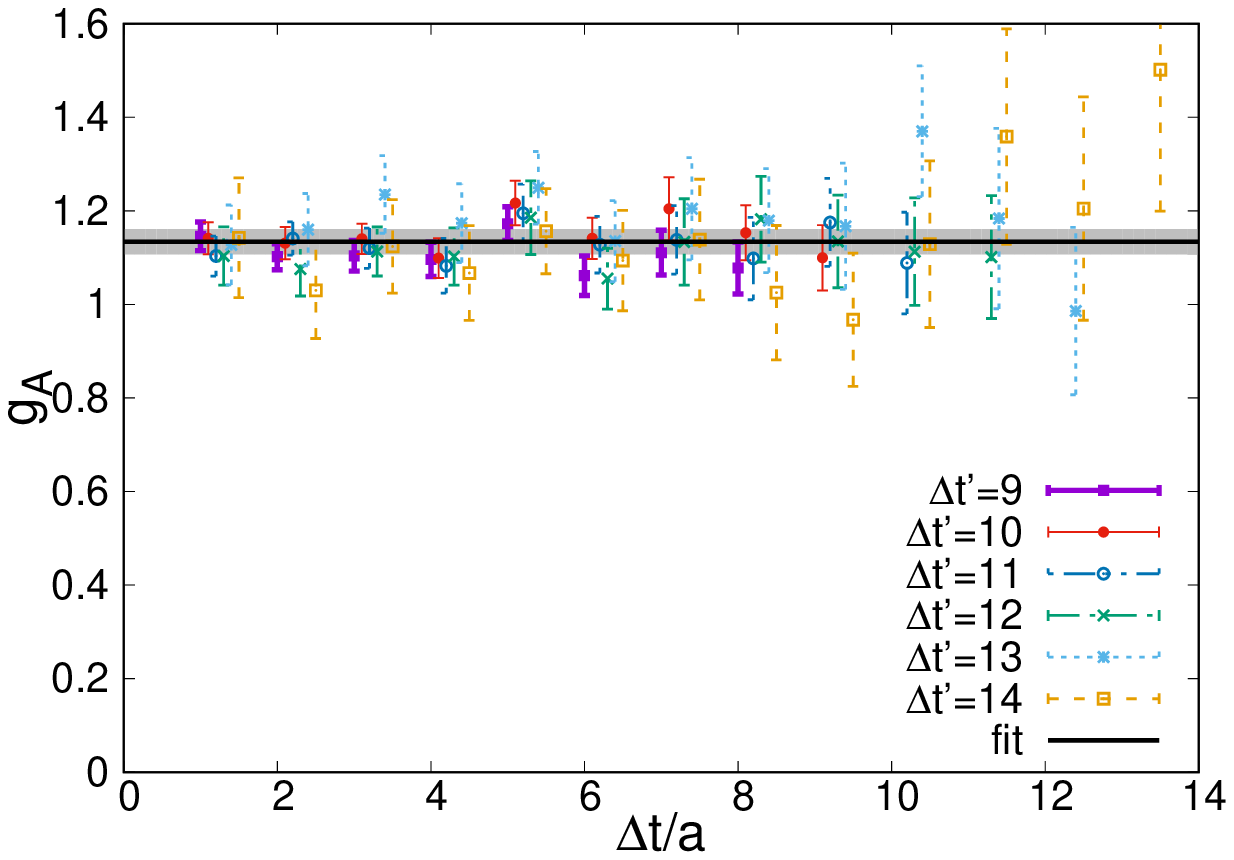}
\includegraphics[width=0.48\textwidth,clip]{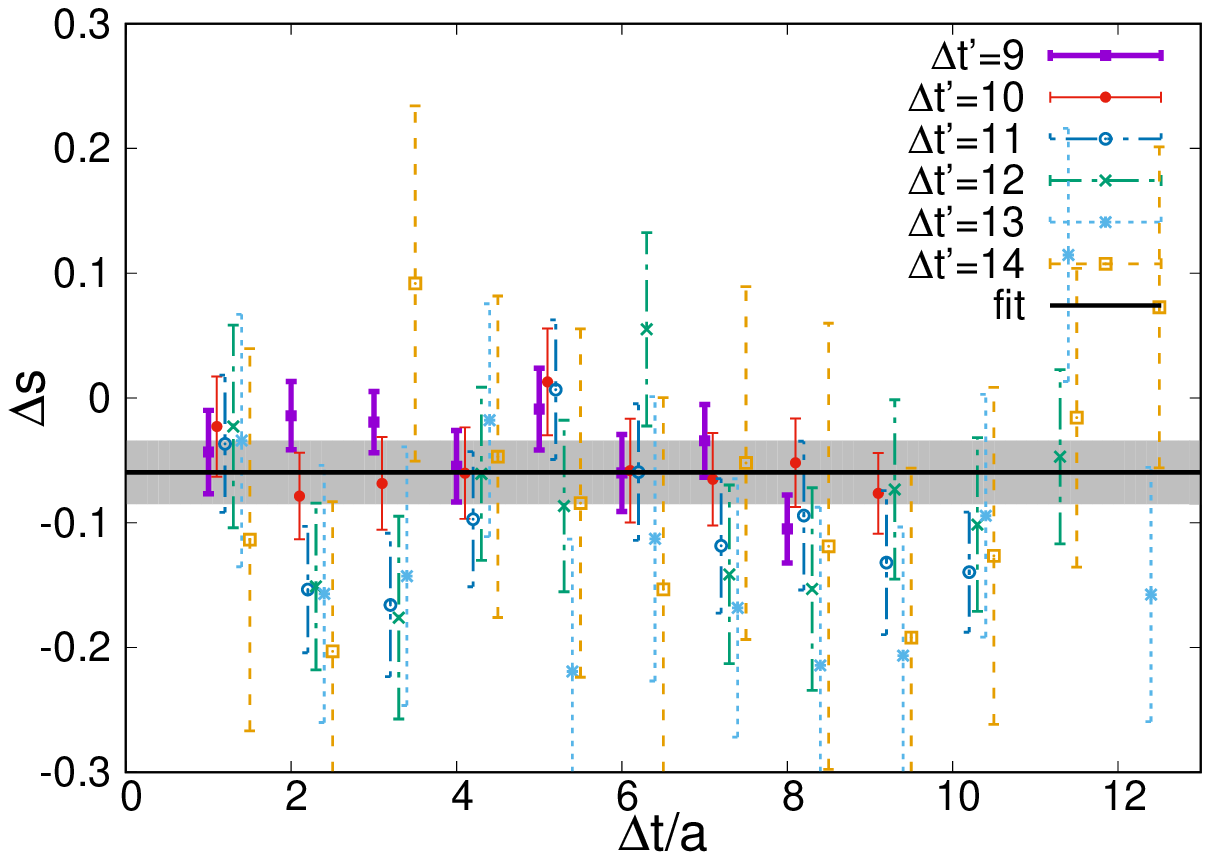}
\vspace{1mm}
\includegraphics[width=0.48\textwidth,clip]{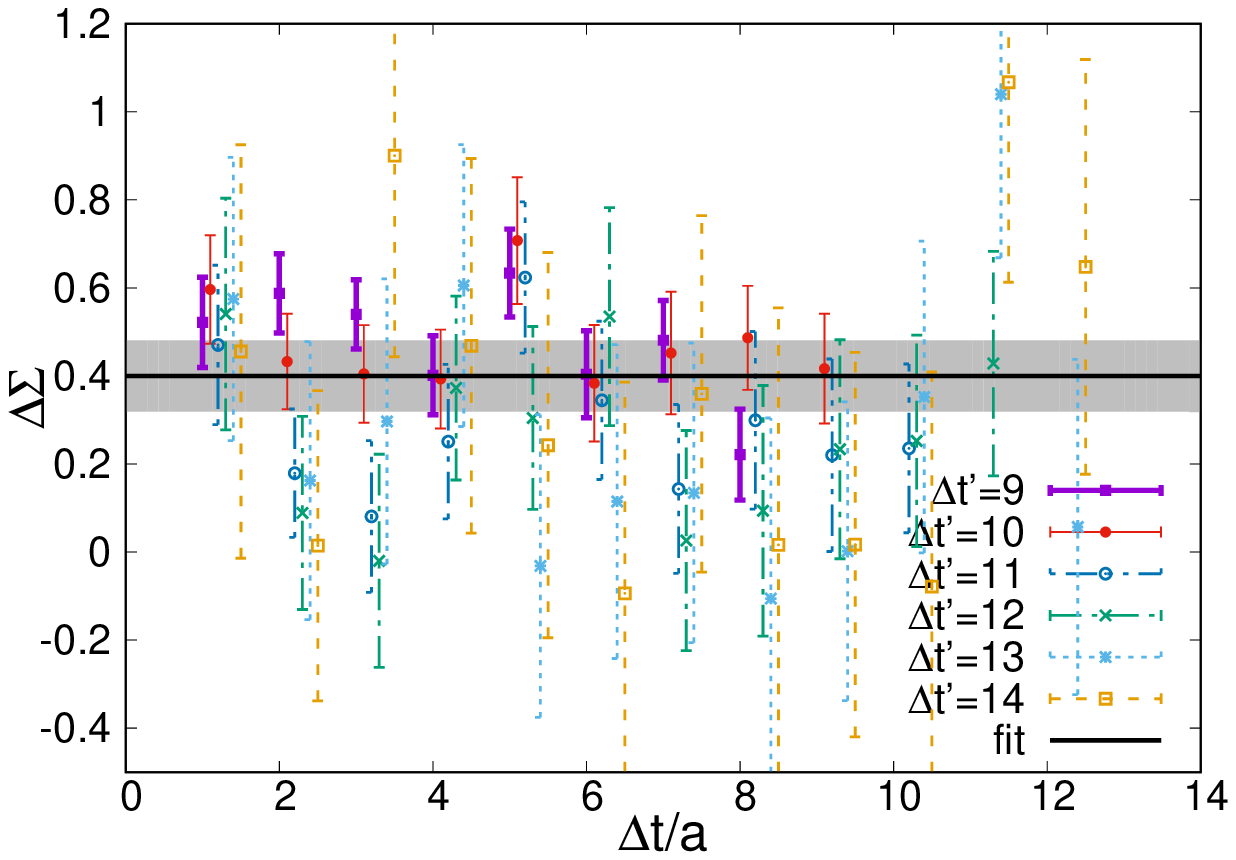}
\caption{\label{fig:axial_charge_fit}
Effective values of axial charges at $m_{ud}\!=\!0.015$ calculated from $R_A(\Delta t,\Delta t^\prime)$. 
The top-left, top-right and bottom panels show data of $g_A$, $\Delta s$ and $\Delta \Sigma$, respectively.
Data with different $\Delta t^\prime$'s are plotted by different symbols as a function of $\Delta t$. 
The horizontal line and the grey band show the fitted value and statistical error of each charge.
}
\end{figure}

For the disconnected diagrams calculated in this study, we observe that a large part of their statistical error comes from a piece which is the product of the low-mode component of the nucleon propagator and the high-mode part of the quark loop.
The statistical accuracy is therefore improved by applying the LMA to the nucleon propagator. 
Figure~\ref{fig:discerror} demonstrates such improvement by taking $\delta s$
as an example. 
We observe that the statistical error scales as $\propto 1/\sqrt{N_{\rm LMA}}$ up to $N_{\rm LMA} \lesssim 200$, beyond which the correlation among different source points is not small.

\section{\label{sec:axial_analysis}Axial charges}

\begin{table}[b]
\begin{center}
\begin{tabular}{l|lccc}
\hline \hline
&$m_{ud}$ & $m_{\pi}$ (MeV) & Charge & $\chi^2$/d.o.f. \\
\hline
$g_A$
&0.050 & 540 & 1.176(27) & 0.23 \\
&0.035 & 450 & 1.152(28) & 0.90 \\
&0.025 & 380 & 1.149(17) & 0.89 \\
&0.015 & 290 & 1.134(27) & 0.58 \\
\hline
$\Delta s$ 
&0.050 & 540 & -0.035(20) & 1.04 \\
&0.035 & 450 & -0.039(25) & 0.37 \\
&0.025 & 380 & -0.026(18) & 0.49 \\
&0.015 & 290 & -0.060(25) & 0.86 \\
\hline
$\Delta \Sigma$ 
&0.050 & 540 & 0.551(70) & 0.90 \\
&0.035 & 450 & 0.478(88) & 0.36 \\
&0.025 & 380 & 0.508(61) & 0.53 \\
&0.015 & 290 & 0.400(81) & 1.07 \\
\hline
\end{tabular}
\end{center}
\caption{
Numerical results for axial charges from constant fit to $R_A(\Delta t,\Delta t^\prime)$ at simulation points.
}
\label{table:nucleon_axial_charges}
\end{table}

Since the axial charges $\Delta u$,  $\Delta d$, and $\Delta s$ are separately calculated, we can consider three independent linear combinations of these.
In this study, we mainly present results for the phenomenologically interesting ones, the isovector charge $g_A$ and singlet charge $\Delta \Sigma$.
They are to be compared with the experimental data \cite{ucna}
\begin{equation}
g_A 
=
1.2783 \pm 0.0022.
\label{eq:gaexp}
\end{equation}
and \cite{compass2}
\begin{equation}
\Delta \Sigma
= 
\Delta u + \Delta d + \Delta s
\in
[0.26, 0.36]
,
\label{eq:totalaxialchargeexp}
\end{equation}
where $\Delta \Sigma$ is given at the renormalization scale $\mu^2 = 3$ GeV$^2$.
We also calculate the strange quark contribution $\Delta s$, which has not been accurately known by experiments.

In Fig.~\ref{fig:axial_charge_fit}, we plot the effective values of these charges calculated from $R_A(\Delta t,\Delta t^\prime)$ at $m_{ud} = 0.015$ as a function of $\Delta t$.
We observe reasonably good plateaux with $\Delta t' \gtrsim 9$ and $\Delta t \in [ 4 , \Delta t' -4]$.
The Gaussian smearing (\ref{eq:gaussian_smearing}) works well in suppressing the excited state contamination also at other $m_{ud}$'s.
We determine $g_A$, $\Delta s$, and $\Delta \Sigma$ by a constant fit in $\Delta t$ and $\Delta t^\prime$.
This fit yields $\chi^2$/d.o.f. $\lesssim 1$.
Numerical results are summarized in Table \ref{table:nucleon_axial_charges}.
In contrast to $g_A$, $\Delta s$ and $\Delta \Sigma$ have larger statistical uncertainty, which dominantly comes from the high-mode contribution to the disconnected quark loop.

\begin{figure}[bp]
\centering
\includegraphics[width=0.48\textwidth,clip]{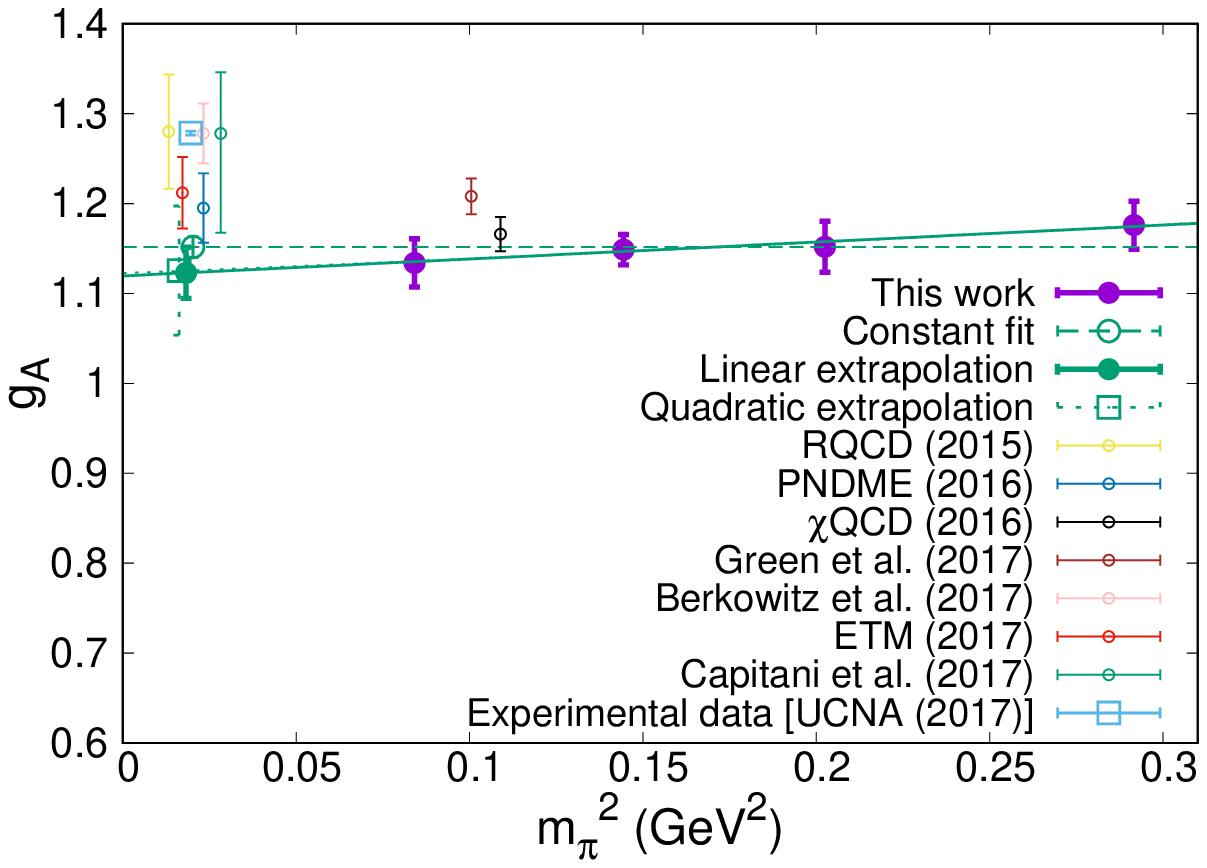}
\includegraphics[width=0.48\textwidth,clip]{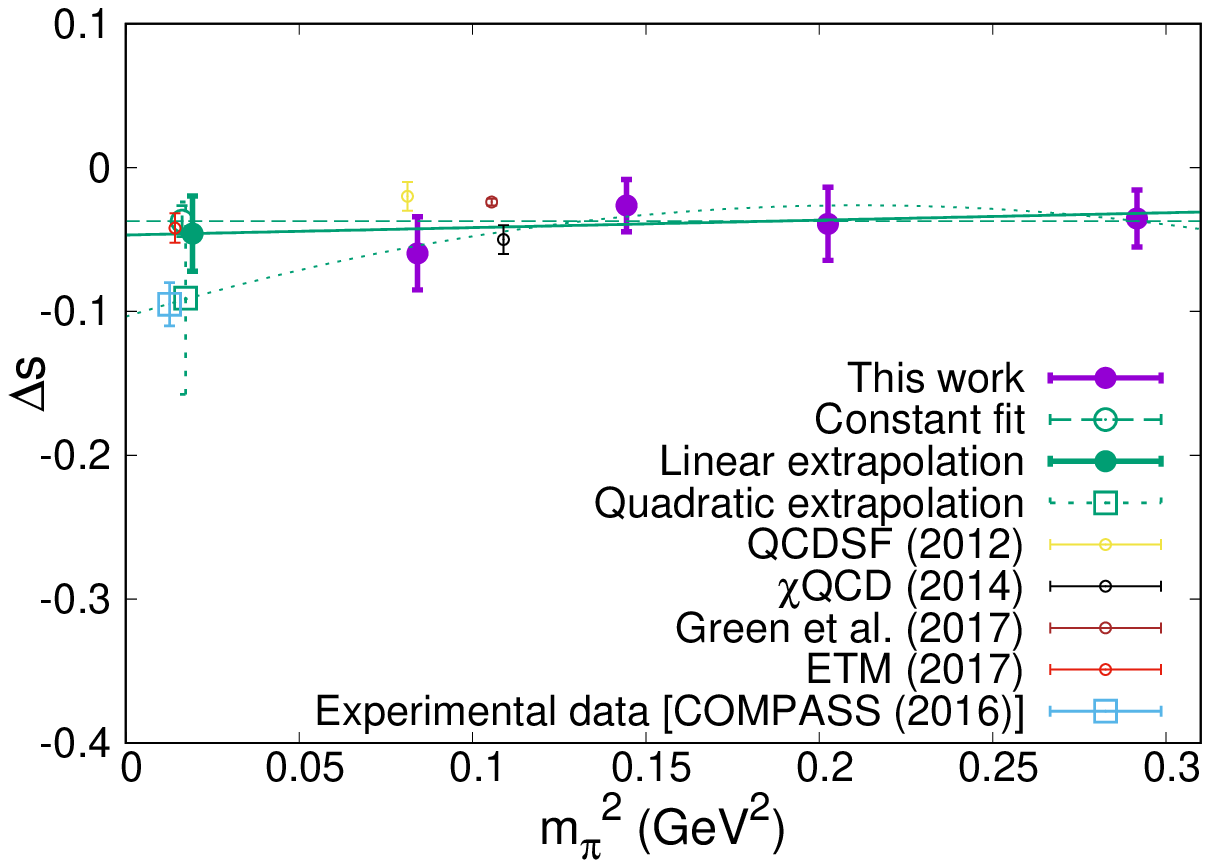}
\vspace{1mm}
\includegraphics[width=0.48\textwidth,clip]{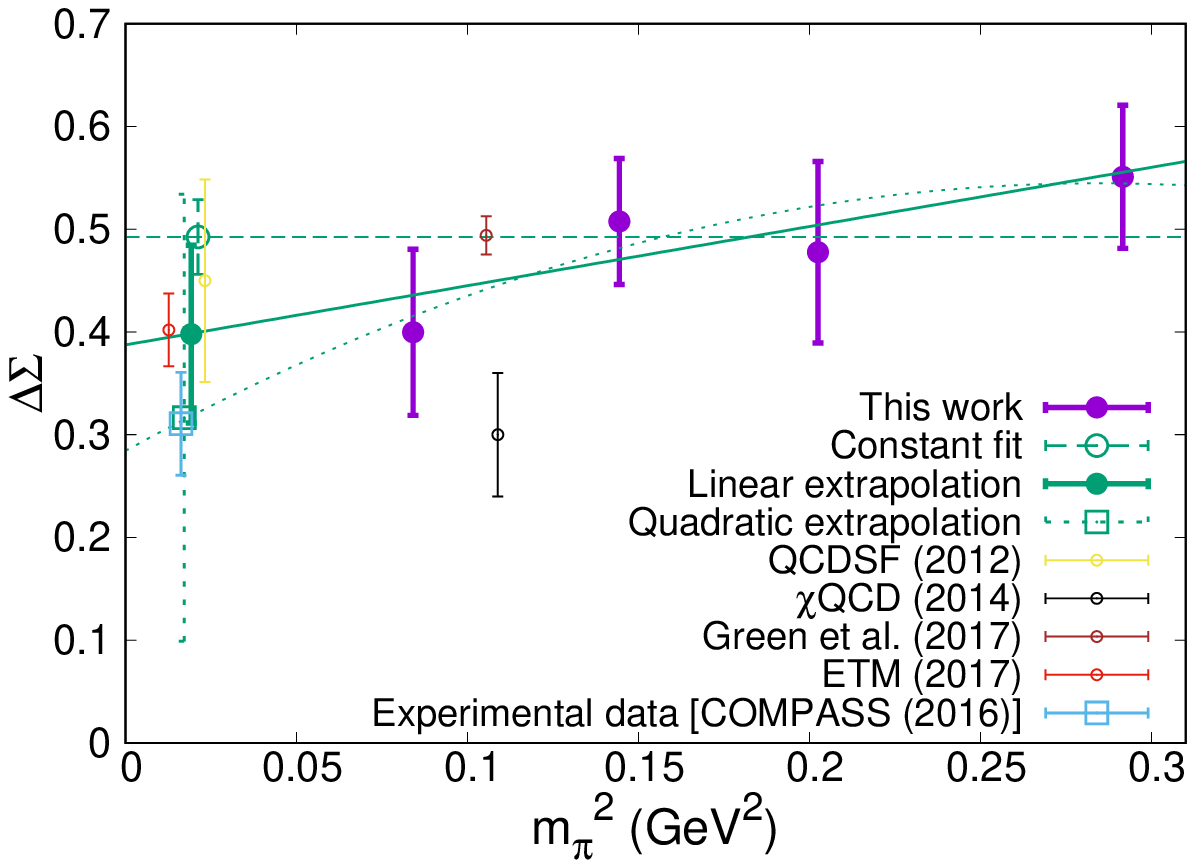}
\caption{\label{fig:axial_charge_extrapolation}
Chiral extrapolation of axial charges.
The top-left, top-right and bottom panels are for 
$g_A$, $\Delta s$, and $\Delta \Sigma$, respectively.
Filled circles show our results, whereas the blue open squares are the experimental values \cite{ucna,compass2}.
The dashed, solid and dotted lines show the constant, linear and quadratic fits, respectively.
We also plot results from other recent studies by small open circles (Refs.~\cite{rqcdisovector,pndmeisovector,chiqcdisovector,etmaxial2017,greenaxial,Berkowitzaxial,capitani} for $g_A$, 
Refs.~\cite{qcdsfprotonspin,chiqcdprotonspin,greenaxial,etmprotonspin} for $\Delta s$, and \cite{qcdsfprotonspin,chiqcdprotonspin,greenaxial,etmprotonspin} $\Delta \Sigma$). 
We note that we changed the renormalization scale of the data of Ref. \cite{compass2} from $\mu^2 =3$ GeV$^2$ to $\mu = 2$ GeV according to the two-loop level renormalization \cite{kodaira}.
}
\end{figure}

\begin{table}[tb]
\begin{center}
\begin{tabular}{l|l|c|cccc}
\hline \hline
& Fitting form & Charge at $m_\pi = 135$ MeV & $c_0$ & $c_1$ & $c_2$ & $\chi^2$/d.o.f. \\
\hline
$g_A $ 
& Constant & 1.152(12) & 1.15(1) & $-$ & $-$ & 0.43 \\
& Linear & 1.123(28) & 1.12(3) & 0.19(17) & $-$ & 0.03 \\
& Quadratic & 1.125(72) & 1.12(9) & 0.2(1.0) & 0.1(2.6) & 0.06 \\
& Eq. (\ref{eq:chiral_perturbation_ga}) & 0.960(10) & 0.88(1) & $-$ & $-$ & 2.19 \\
& Eq. (\ref{eq:chiral_perturbation_ga_improved}) & 0.986(15) & 0.90(1) & $-$ & -1.04(45) & 0.83 \\
\hline
$\Delta s $ 
& Constant & -0.037(11) & -0.037(11) & $-$ & $-$ & 0.38 \\
& Linear & -0.046(26) & -0.047(29) & 0.05(14) & $-$ & 0.51 \\
& Quadratic & -0.091(67) & -0.104(83) & 0.73(94) & -1.7(2.4) & 0.48 \\
\hline
$\Delta \Sigma $ 
& Constant & 0.492(36) & 0.492(36) & $-$ & $-$ & 0.70 \\
& Linear & 0.398(86) & 0.387(94) & 0.58(48) & $-$ & 0.33 \\
& Quadratic & 0.317(218) & 0.284(270) & 1.83(3.12) & -3.2(7.9) & 0.49 \\
\hline
\end{tabular}
\end{center}
\caption{
Numerical results of chiral extrapolations of axial charges.
}
\label{table:nucleon_axial_charges_extrapolation}
\end{table}

Figure~\ref{fig:axial_charge_extrapolation} shows the results for $g_A$, $\Delta s$, and $\Delta \Sigma$ as a function of $m_\pi^2$.
We observe mild $m_\pi^2$ dependence of all the axial charges.
For the extrapolation to the physical point $m_{\pi , {\rm phys}}$, we test the constant, linear, and quadratic fits,
\begin{equation}
g 
=
c_0 + c_1 m_\pi^2 + c_2 m_\pi^4
,
\label{eq:polynomial_extrapolation}
\end{equation}
where $g$ represents the charge to be fitted.
Numerical results of these  polynomial fits and extrapolated values are summarized in Table \ref{table:nucleon_axial_charges_extrapolation}.
Due to the mild $m_{\pi}^2$ dependence, coefficients $c_1$ and $c_2$ are consistent with zero.
While the constant fit gives acceptable values of $\chi^2/{\rm d.o.f.}\!\sim\!0.4$\,--\,0.7, we do not rule out the $m_\pi^2$ dependence and conservatively employ the linear fit.
The systematic uncertainty due to the choice of the fitting form is estimated from the difference between the linear and constant fits.
Here we do not use the quadratic fit,
which involves an additional ill-determined fit parameter $c_2$. 
In Sections~\ref{sec:scalar_analysis} and \ref{sec:tensor_analysis},
the same manner is used for the scalar and tensor charges,
for which the linear coefficient $c_1$ is consistent with zero
within 2~$\sigma$. 
Since we simulate the single lattice spacing, our results are subject to discretization errors, which are estimated as $O((a\Lambda_{\rm QCD})^2)\!\approx\!8$\,\% by taking $\Lambda_{\rm QCD}\!\approx\!500$~MeV.
Our results for the axial charges are 
\begin{eqnarray}
g_A 
&=&
1.123 (28)_{\rm stat} (29)_{\chi} (90)_{a \neq 0}
,
\label{eq:our_g_a}
\\
\Delta s
&=&
-0.046 (26)_{\rm stat} (9)_{\chi} 
,
\\
\Delta \Sigma
&=&
0.398 (86)_{\rm stat} (94)_{\chi} (32)_{a \neq 0}
.
\end{eqnarray}
The first error is the statistical error. 
The second and third errors are the systematic ones due to the chiral extrapolation and finite lattice spacing, respectively.
Here we neglect the discretization error for $\Delta s$, which is much smaller than its total uncertainty and the systematic error due to the extrapolation.
Our result for $g_A$ is consistent with those of previous lattice studies~\cite{rqcdisovector,chiqcdisovector,pndmeisovector,greenaxial,etmaxial2017,Berkowitzaxial,capitani} and with the experimental value (\ref{eq:gaexp}) within 8\% discretization error.

\begin{figure}[bp]
\centering
\includegraphics[width=0.48\textwidth,clip]{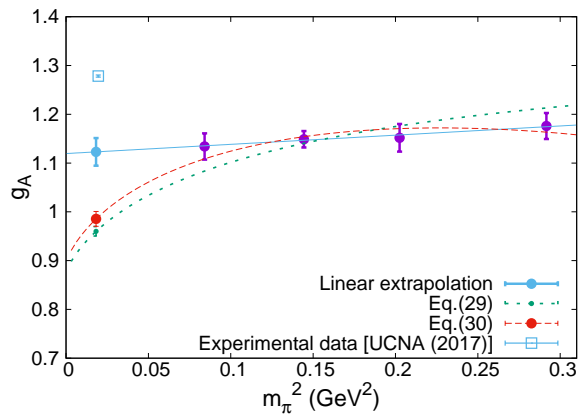}
\caption{\label{fig:extrapolation_isovector_axial_charge}
Chiral extrapolation of $g_A$ using ChPT-based fitting forms of Eqs.~(\ref{eq:chiral_perturbation_ga}) (dotted line) and (\ref{eq:chiral_perturbation_ga_improved}) (dashed line).
The solid line shows the linear fit leading to our result~(\ref{eq:our_g_a}).
}
\end{figure}

The isovector charge $g_A$ has been calculated within one-loop ChPT~\cite{Bijnens,Jenkins,detmold}.
We also test extrapolations based on ChPT for $g_A$ in order to check the consistency between our lattice data and the nonanalytic chiral behavior predicted by this theory.
In this study, we employ the one-loop formula in Ref.~\cite{detmold} 
\begin{equation}
g_A
=
c_0 \Biggl[
1+\frac{m_\pi^2}{ (4\pi f_\pi )^2}
(1+2 
c_0^2
) \ln \Bigl( \frac{\mu^2+m_\pi^2}{m_\pi^2 } \Bigr)
\Biggr]
+ \frac{m_\pi^2}{m_\pi^2 + (5\, {\rm GeV})^2}
\Biggl[
\frac{5}{3} - c_0 \Biggl(
1+ \mu^2 \frac{1+2 
c_0^2
}{ (4\pi f_\pi )^2}
\Biggr)
\Biggr]
,
\label{eq:chiral_perturbation_ga}
\end{equation}
where we employ the experimental value $f_\pi = 93$~MeV by ignoring higher order corrections.
A parameter $\mu$ = 550 MeV is introduced by the authors of Ref.~\cite{detmold} to suppress the rapid variation of the logarithmic term away from the chiral limit.
The last term in the curly bracket is also introduced in Ref. \cite{detmold} so that $g_A$ converges to the quark model estimate $\frac{5}{3}$ in the heavy quark limit.
This term is, however, not large at $m_\pi\!\ll5$~GeV, and has small influence in the following discussion.
Here $c_0$ is the only fit parameter.

Numerical results of the ChPT-based extrapolation is also listed in Table \ref{table:nucleon_axial_charges_extrapolation}.
As shown in Fig. \ref{fig:extrapolation_isovector_axial_charge}, the one-loop formula (\ref{eq:chiral_perturbation_ga}) fails to describe our data with $\chi^2/{\rm d.o.f.}\!\gtrsim\!2$.
We then include a higher order analytic term
\begin{equation}
g_A
=
\mbox{``right-hand side of Eq.~(\ref{eq:chiral_perturbation_ga})''}
+ c_2 m_\pi^4
\, .
\label{eq:chiral_perturbation_ga_improved}
\end{equation}
While this fit obtains reasonable $\chi^2/{\rm d.o.f}\!\sim\!0.8$, the extrapolated value is well below the experimental value (see Table \ref{table:nucleon_axial_charges_extrapolation} and Fig. \ref{fig:extrapolation_isovector_axial_charge}).
It is difficult to describe both the lattice and experimental data within one-loop ChPT probably because of significant higher order corrections in our simulation region of $m_\pi$ \cite{bernard}.
We therefore do not take account of these ChPT fits in our error estimate for $g_A$ in Eq.~(\ref{eq:our_g_a}).

\begin{figure}[bp]
\centering
\includegraphics[width=0.48\textwidth,clip]{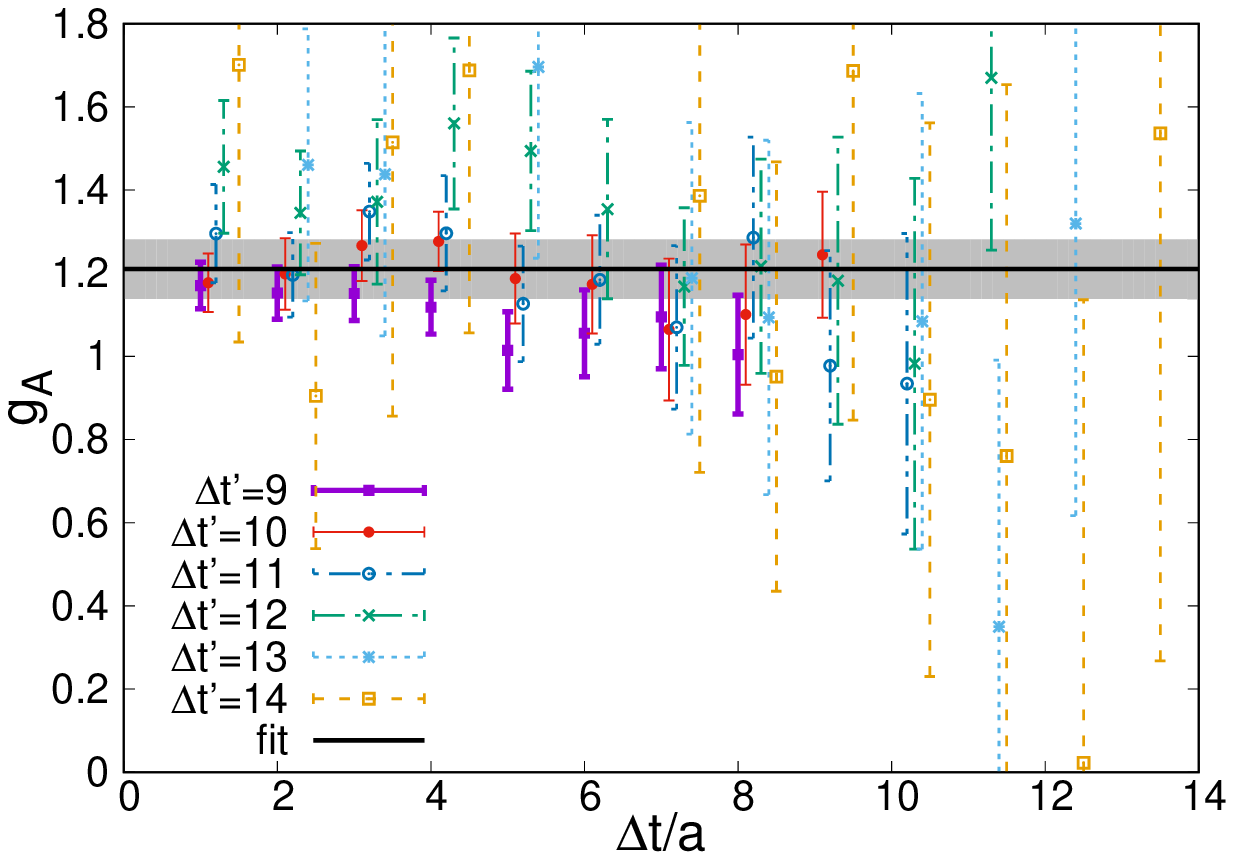}
\includegraphics[width=0.48\textwidth,clip]{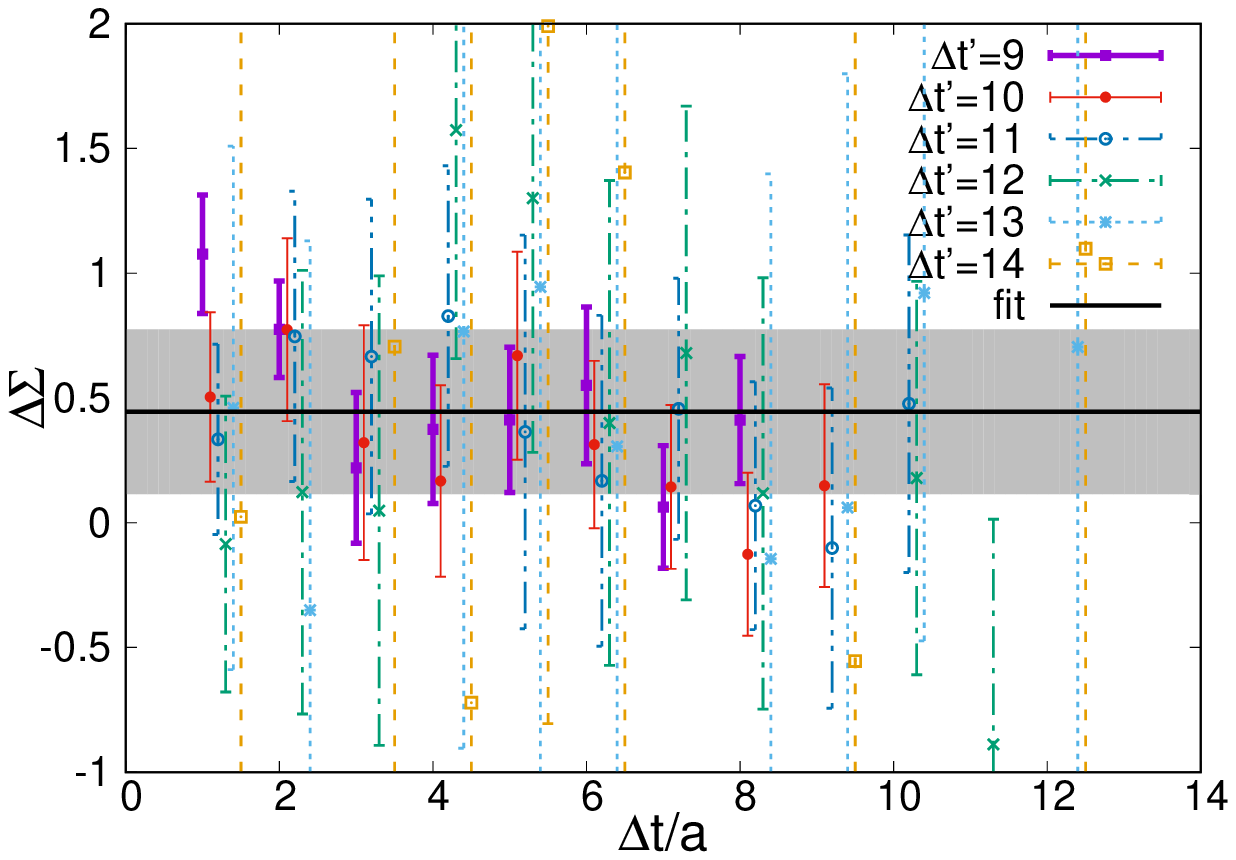}
\caption{\label{fig:Q=1_axial_charge}
Effective values of $g_A$ (left panel) and $\Delta \Sigma$ (right panel) 
in nontrivial topological sector with $Q=1$ at $m_{ud} = 0.015$.
}
\end{figure}

The singlet axial charge $\Delta \Sigma$ has the same quantum number as the topological charge.
The effect of the fixed topology in our simulation may be important, although it is suppressed by the inverse space-time volume $1/V$.
We examine the effect by simulating the nontrivial topological sector $Q=1$ on a $16^3 \times 48$ lattice at $m_{ud}= 0.015$.
Effective values of $g_A$ and $\Delta \Sigma$ are plotted in Fig.~\ref{fig:Q=1_axial_charge}.
The constant fit in $\Delta t$ and $\Delta t^\prime$ yields $g_A = 1.210 (72)$ and $\Delta \Sigma = 0.44(33)$. 
While the statistical accuracy of $\Delta \Sigma$ is not high, agreement with those for $Q=0$ in Table~\ref{table:nucleon_axial_charges} suggests that the fixed topology effect is not large.

The extrapolated value of $\Delta \Sigma$ and those at small $m_{ud}\!\leq\!0.025$ are systematically smaller than unity.
We also note that our result for $\Delta s$ is consistent with the experimental value $\Delta s_{\rm exp}  \in [-0.11 , -0.08]$~\cite{compass2}: namely the spin contribution of strange sea quarks is not large.
This is consistent with the proton spin puzzle stating that the nucleon spin is not saturated by the quark spin contribution.

\section{\label{sec:scalar_analysis}Scalar charges}

\begin{figure}[bp]
\centering
\includegraphics[width=0.48\textwidth,clip]{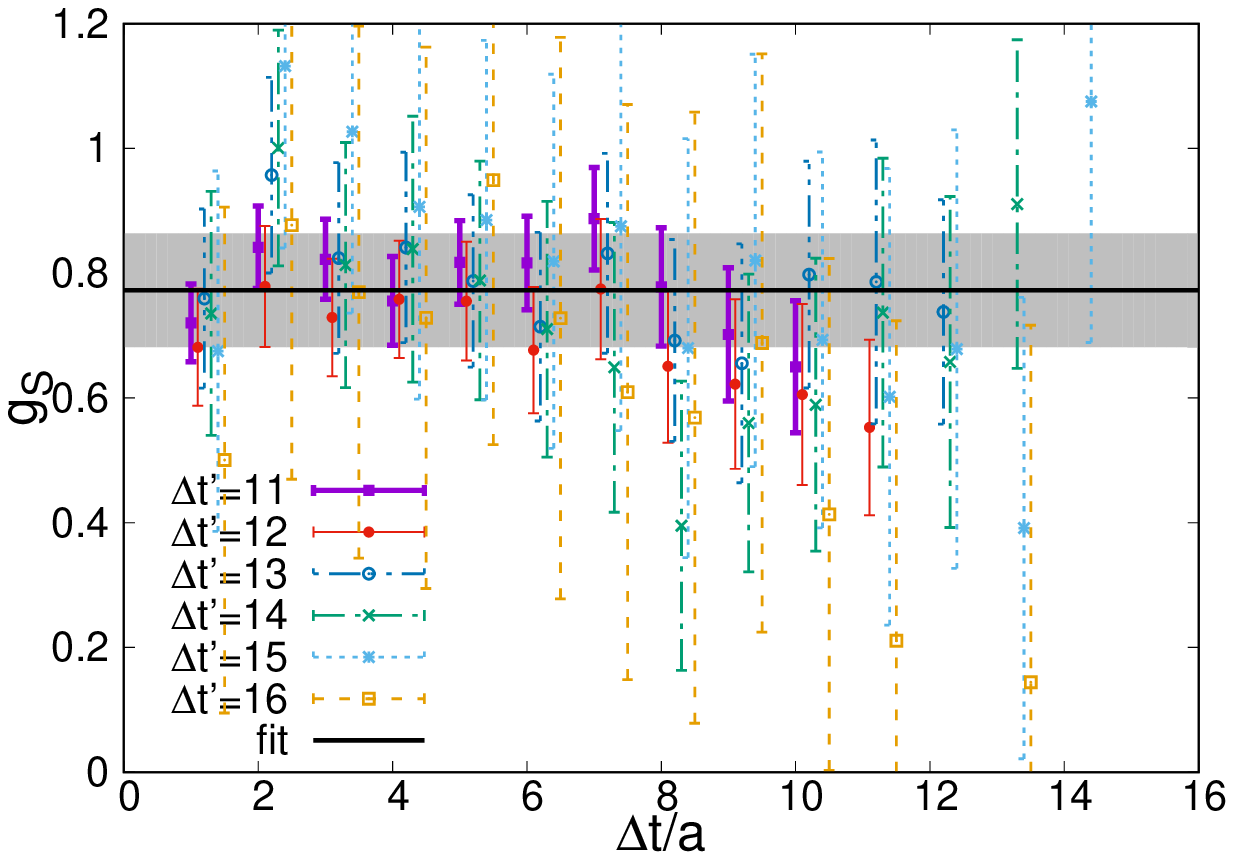}
\includegraphics[width=0.48\textwidth,clip]{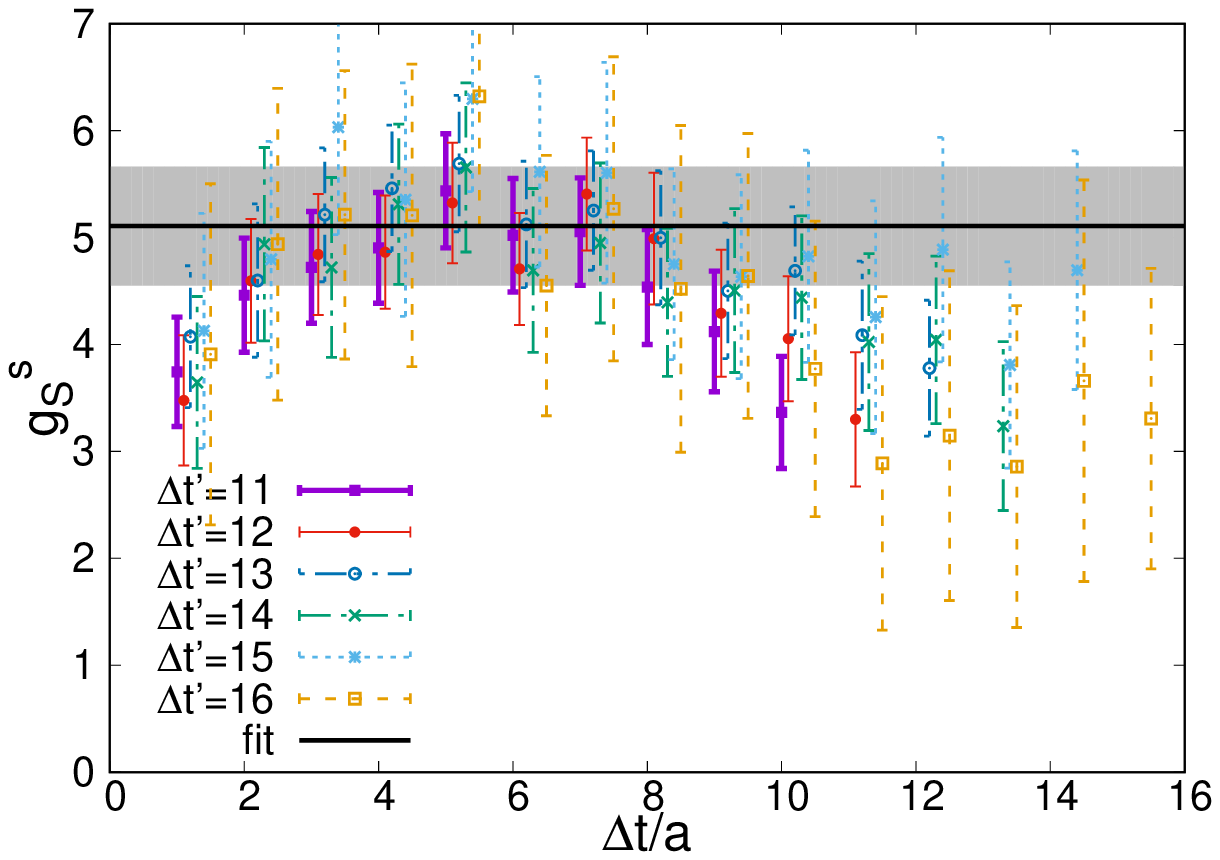}
\vspace{1mm}
\includegraphics[width=0.48\textwidth,clip]{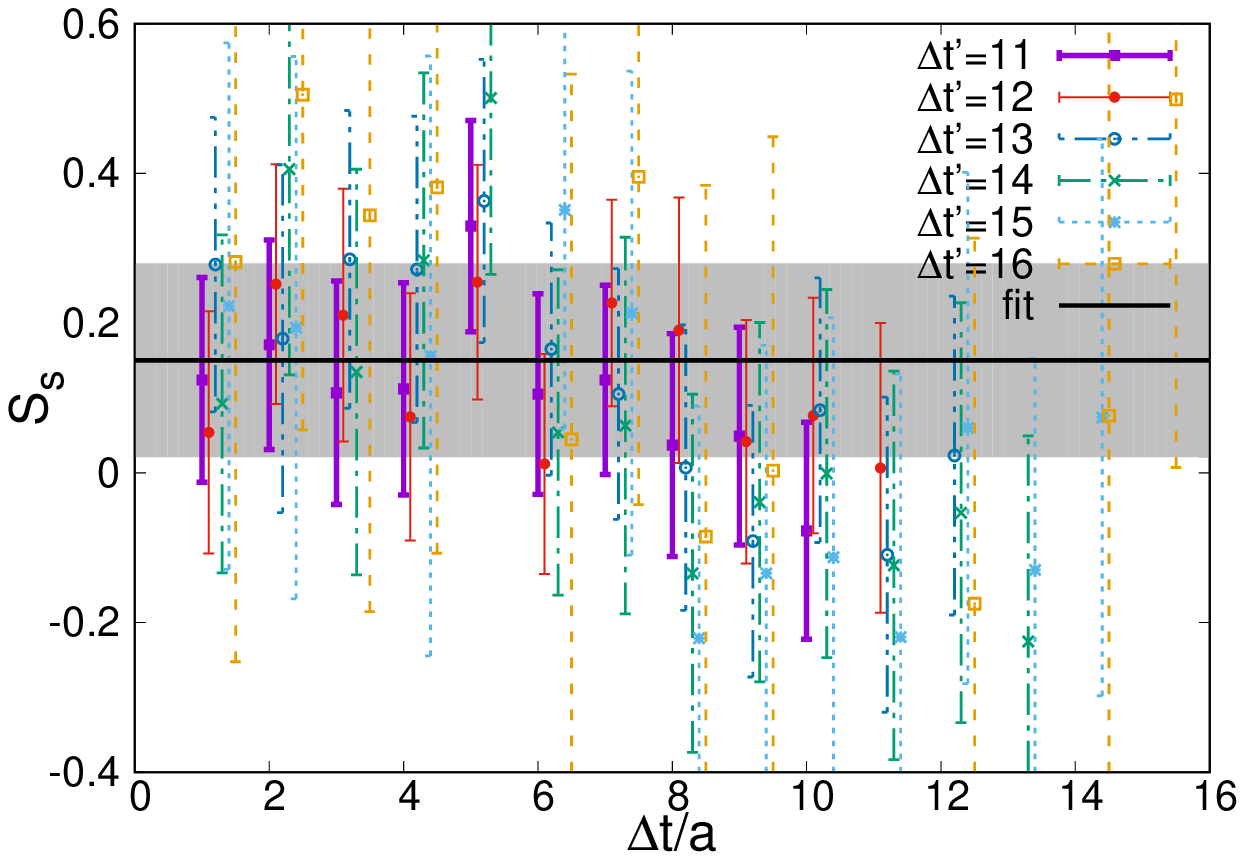}
\includegraphics[width=0.48\textwidth,clip]{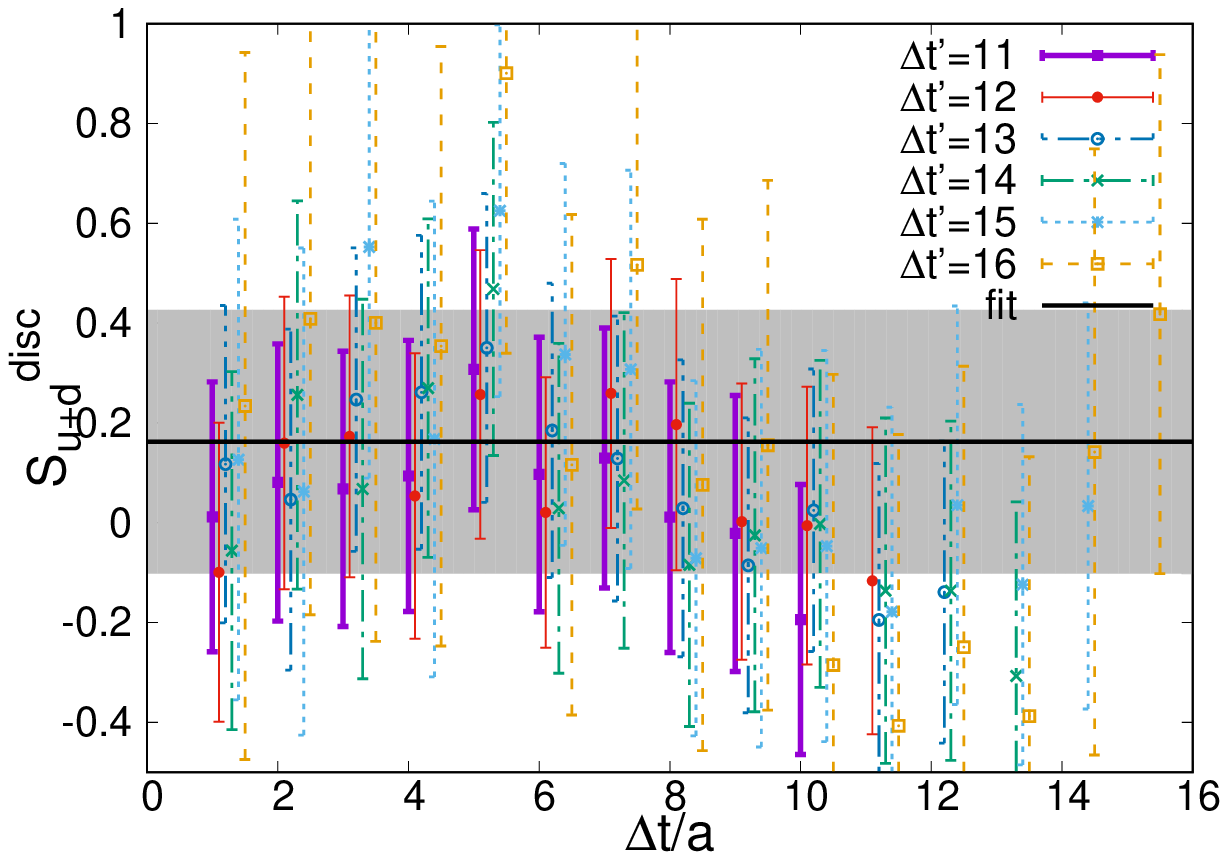}
\caption{\label{fig:scalar_charge_fit}
Effective values of scalar charges at $m_{ud}\!=\!0.015$ as a function of $\Delta t$.
We plot data of $g_S$ and $S_s$ in the top- and bottom-left panels, whereas the top- and bottom-right panels are for $g_S^s$ and its disconnected contribution $S_{u+d}^{\rm disc}$.
}
\end{figure}

For the scalar charges (\ref{eq:scalar_charge}), we consider isovector and isoscalar combinations,
\begin{eqnarray}
g_S
\equiv
\frac{1}{2m_N}
\langle p | \bar u u - \bar d d | p \rangle
= S_u - S_d 
, \hspace{5mm}
g_S^s
\equiv
\frac{1}{2m_N}
\langle p | \bar u u + \bar d d | p \rangle
= S_u + S_d,
\end{eqnarray}
and the strange quark contribution $S_s$.
Note that $g_S^s$ and $S_s$ are related to the pion-nucleon sigma term and the strange quark content as 
\begin{equation}
\sigma_{\pi N} = m_{ud} g_S^s,
\hspace{5mm}
\sigma_s = m_s S_s.
\label{eq:sigma_term}
\end{equation}
We also consider the disconnected contribution $S_{u+d}^{\rm disc}$, namely the second term in Eq.~(\ref{eq:3ptfunction}) to test the Okubo-Zweig-Iizuka (OZI) rule.

\begin{table}[bp]
\begin{center}
\begin{tabular}{l|lccc}
\hline \hline
& $m_{ud}$ & $m_{\pi}$ (MeV) & Charge & $\chi^2$/d.o.f. \\
\hline
$g_S$ 
&0.050 & 540 & 0.824(43) & 0.42 \\
&0.035 & 450 & 0.854(67) & 0.11 \\
&0.025 & 380 & 0.898(49) & 0.57 \\
&0.015 & 290 & 0.773(91) & 0.30 \\
\hline
$g_S^s$ 
&0.050 & 540 & 4.30(35) & 0.52 \\
&0.035 & 450 & 4.40(53) & 0.54 \\
&0.025 & 380 & 5.35(52) & 0.29 \\
&0.015 & 290 & 5.11(56) & 0.47 \\
\hline
$S_{u+d}^{\rm disc}$ 
&0.050 & 540 & 0.34(16) & 0.55 \\
&0.035 & 450 & 0.29(20) & 0.80 \\
&0.025 & 380 & 0.44(25) & 0.26 \\
&0.015 & 290 & 0.16(26) & 0.39 \\
\hline
$S_s$ 
&0.050 & 540 & 0.26(13) & 0.63 \\
&0.035 & 450 & 0.28(13) & 1.07 \\
&0.025 & 380 & 0.25(17) & 0.45 \\
&0.015 & 290 & 0.15(13) & 0.83 \\
\hline
\end{tabular}
\end{center}
\caption{
Numerical results for scalar charges from constant fit to $R_S(\Delta t,\Delta t^\prime)$ at simulation points.
}
\label{table:nucleon_scalar_charges}
\end{table}

We extract $g_S$, $g_S^s$,  $S_s$ and $S_{u+d}^{\rm disc}$ at each simulation point in a way similar to that for the axial charges.
Figure \ref{fig:scalar_charge_fit} shows the effective values of the scalar charges and their constant fits at $m_{ud}=0.015$.
Numerical results are summarized in Table \ref{table:nucleon_scalar_charges}.
The reasonable plateaux that we observe also at other $m_{ud}$'s lead to $\chi^2/{\rm d.o.f.}\!<\!1.5$ for the constant fit in $\Delta t$ and $\Delta t^\prime$.
Although $g_S^s$ contains the noisy disconnected contribution, it is reasonably dominated by the connected contribution and, hence, is determined with an accuracy of 10\,\%.

On the other hand, $\sigma_s$ is a purely disconnected contribution, and our results are consistent with zero.
We confirm a good agreement with our previous results in Refs.~\cite{takeda,ohki}.
Their statistical uncertainties are also comparable, while this study employs the TSM to average the disconnected diagram over more source points.
This is because the statistical error dominantly comes from a contribution that consists of the high-mode quark loop and the low-mode nucleon propagator. 
This contribution is difficult to improve by the TSM, which is applied only for the high-mode part of the point-to-all propagators.

\begin{figure}[bp]
\centering
\includegraphics[width=0.48\textwidth,clip]{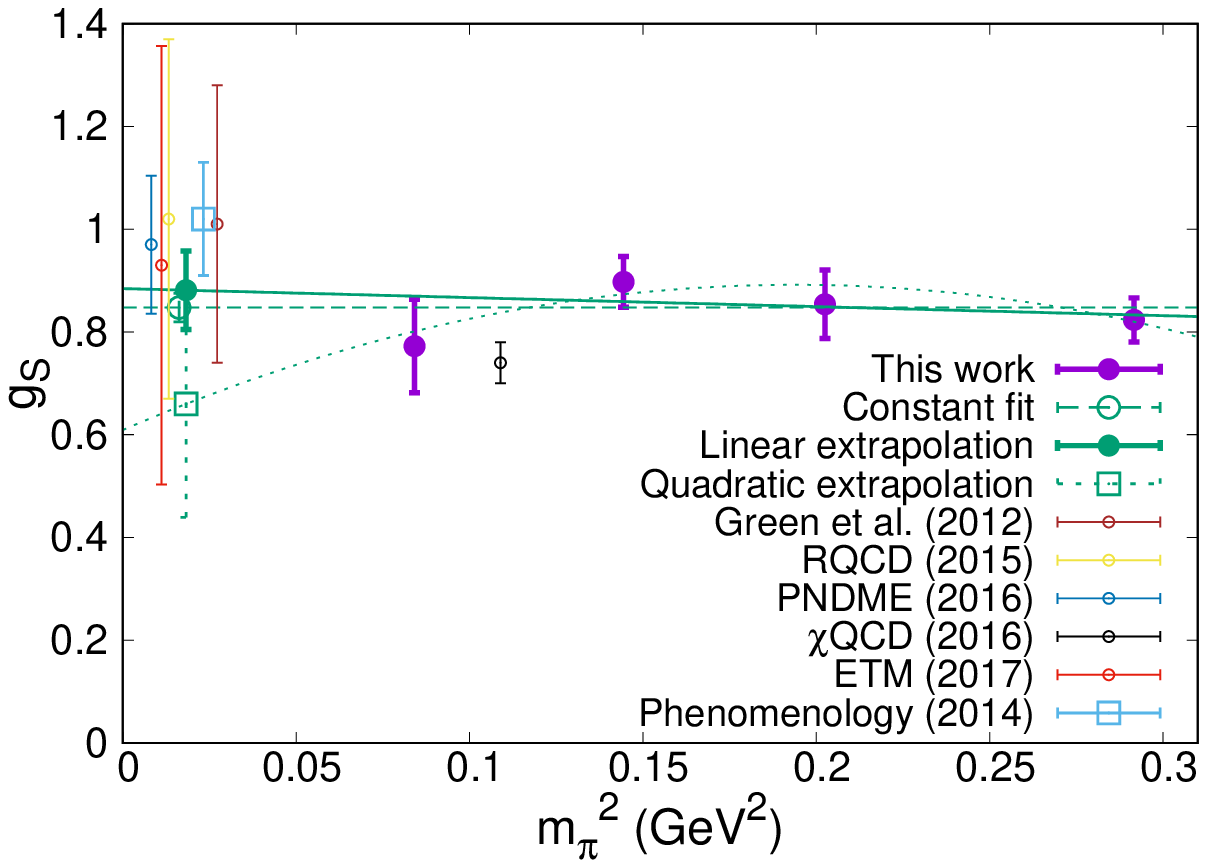}
\includegraphics[width=0.48\textwidth,clip]{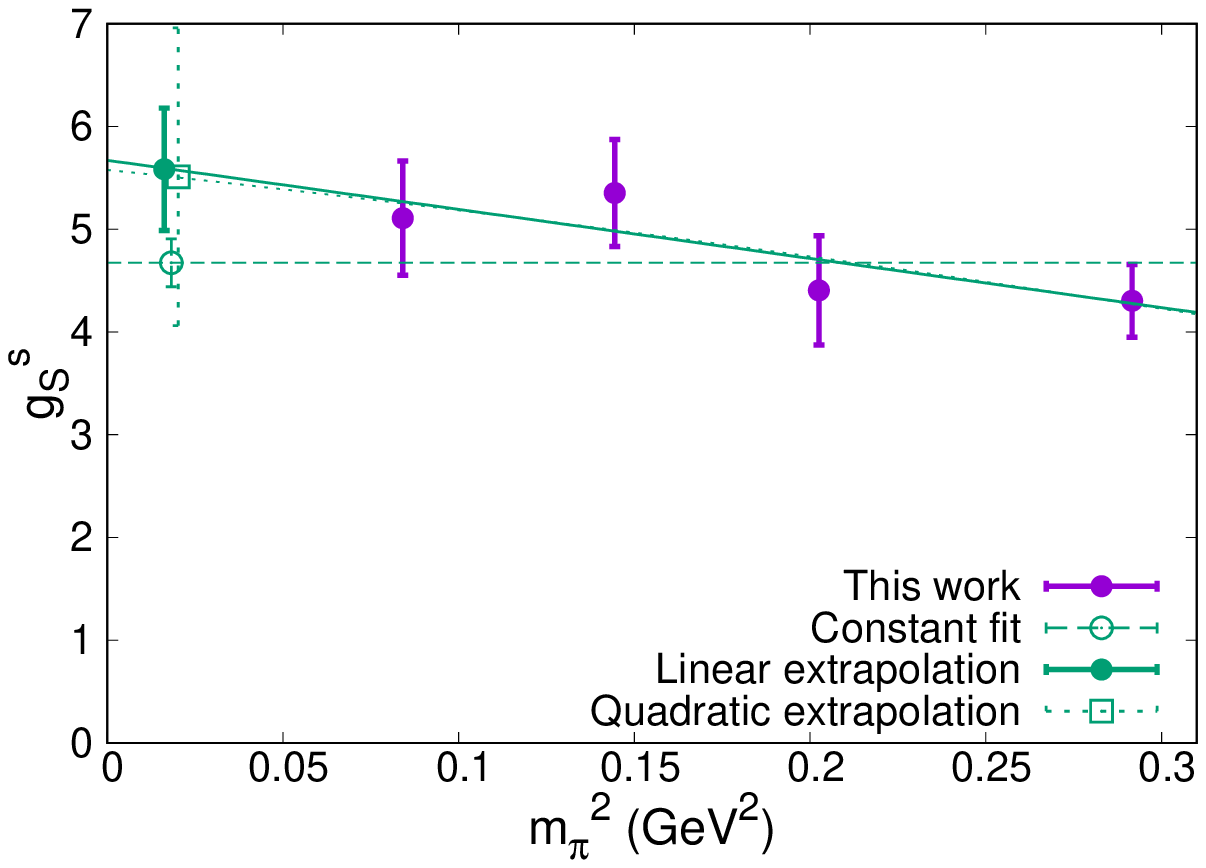}
\vspace{1mm}
\includegraphics[width=0.48\textwidth,clip]{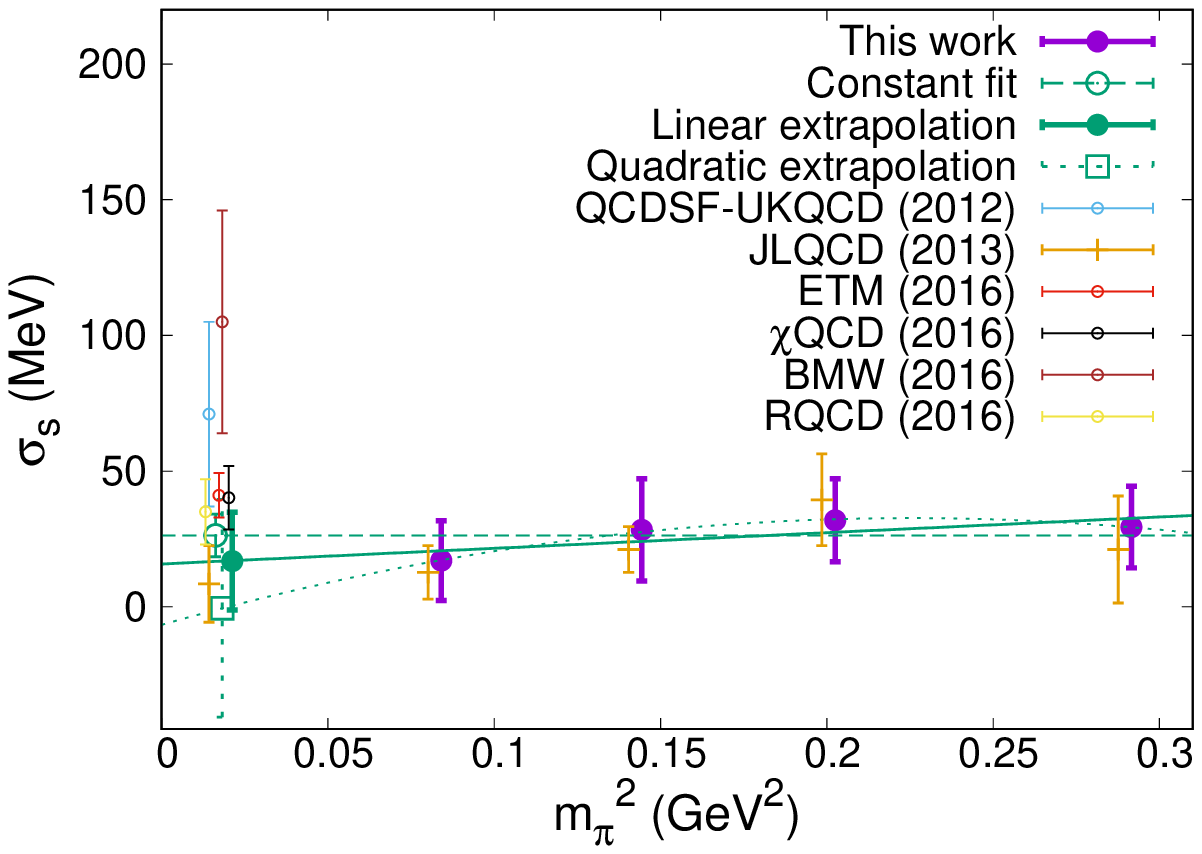}
\caption{\label{fig:scalar_charge_extrapolation}
Chiral extrapolations of scalar charges.
The top-left, top-right and bottom panels show data of $g_S$, $g_S^s$ and $\sigma_s$, respectively.
Filled circles are our results, whereas small open circles are from recent lattice calculations of $g_S$~\cite{rqcdisovector,pndmeisovector,chiqcdisovector,green,etm2017} and $\sigma_s$~\cite{etmsigmaterm,chiqcdsigmaterm,bmwsigmaterm,rqcdsigmaterm,ohki2,qcdsf-ukqcdsigmaterm}.
We also plot a phenomenological estimate of $g_S$~\cite{alonso} by the blue open square.
}
\end{figure}

Our results for the scalar charges are plotted in Fig.~\ref{fig:scalar_charge_extrapolation}.
Here and in the following, we consider $\sigma_s$ instead of $S_s$ for a straightforward comparison with previous lattice calculations.
The scalar charges show mild $m_\pi^2$ dependence in our simulation region of $m_\pi$.
Those for $g_S^s$ and $\sigma_s$ are consistent with our observations in Refs.~\cite{ohki,takeda,ohki2}.
We therefore test the constant, linear, and quadratic fitting form (\ref{eq:polynomial_extrapolation}) to extrapolate them to $m_{\pi,{\rm phys}}$.
Numerical results are summarized in Table~\ref{table:nucleon_scalar_charges_extrapolation}.
As expected from the mild $m_\pi^2$ dependence, the coefficients $c_1$ and $c_2$ are consistent with zero, and the constant fit achieves good values of $\chi^2/\mbox{d.o.f}\!<\!1$.
For $g_S$ and $\sigma_s$, we conservatively employ the linear fit, and the difference in the extrapolated value from the constant fit is treated as the systematic uncertainty due to the choice of the fitting form.

\begin{table}[tb]
\begin{center}
\begin{tabular}{l|l|c|cccc}
\hline \hline
& Fitting form & Charge at $m_\pi = 135$ MeV & $c_0$ &$c_1$ & $c_2$ & $\chi^2$/d.o.f. \\
\hline
$g_S $ 
& Constant & 0.85(3) &0.85(3) & $-$ & $-$ & 0.68 \\
& Linear & 0.88(8) &0.88(8) & -0.18(37) & $-$ & 0.90 \\
& Quadratic & 0.66(22) & 0.61(27) & 2.9(2.9) & -7.5(7.0) & 0.65 \\
\hline
$g_S^s$
& Constant & 4.7(2) & 4.7(2) & $-$ & $-$ & 1.22 \\
& Linear & 5.6(6) & 5.7(6) & -4.8(2.9) & $-$ & 0.46 \\
& Quadratic & 5.5(1.4) & 5.6(1.8) & -4(21) & -3(50) & 0.91 \\
\hline
$S_{u+d}^{\rm disc}$ 
& Constant & 0.32(10) & 0.32(10) & $-$ & $-$ & 0.22 \\
& Linear & 0.24(28) & 0.24(30) & 0.4(1.4) & $-$ & 0.28 \\
& Quadratic & 0.04(65) & -0.02(80) & 3.5(9.0) & -8(22) & 0.45 \\
\hline
$S_s$ 
& Constant & 0.23(7) & 0.23(7) & $-$ & $-$ & 0.19 \\
& Linear & 0.15(16) & 0.14(17) & 0.51(87) & $-$ & 0.12 \\
& Quadratic & 0.00(35) & -0.06(44) & 3.1(5.3) & -7(14) & 0.002 \\
\hline
\end{tabular}
\end{center}
\caption{
Numerical results of polynomial chiral extrapolations of scalar charges.
}
\label{table:nucleon_scalar_charges_extrapolation}
\end{table}

\begin{figure}[tbp]
\centering
\includegraphics[width=0.48\textwidth,clip]{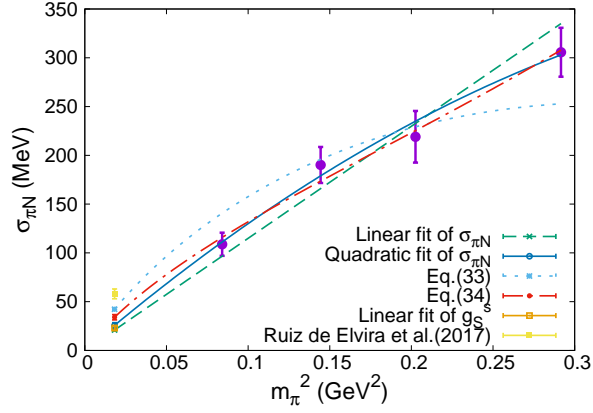}
\caption{\label{fig:extrapolation_sigma_term}
Chiral extrapolations of $\sigma_{\pi N}$. 
The dashed and solid lines show linear and quadratic fits.
The dot-dashed and dot-dot-dashed lines are the extrapolation based on $O(p^3)$ ChPT [Eq. (\ref{eq:chiral_perturbation_sigma_term})] and that including $O(p^4)$ analytic term [Eq. (\ref{eq:chiral_perturbation_sigma_term_2})], respectively.
The value and the uncertainty of $\sigma_{\pi N}$ obtained from the linear fit of $g_S^s$ are also shown at $m_\pi = m_{\pi , {\rm phys}}$.
Note that $\sigma_{\pi N}\!=\!0$ at $m_\pi^2\!=\!0$ from the definition~(\ref{eq:sigma_term}).
}
\end{figure}

Through Eq.~(\ref{eq:sigma_term}), we can convert $g_S^s$ to $\sigma_{\pi N}$, which has an enhanced $m_\pi^2$ dependence due to the overall factor $m_{ud}$.
Since the simple relation $m_\pi^2 \!\propto\! m_{ud}$ receives significant higher order corrections at our simulation points~\cite{Spectrum:Nf3:RG+Ovr:JLQCD}, a comparison between chiral extrapolations of $\sigma_{\pi N}$ and $g_S^s$ provides a check of the stability of the extrapolated value of $\sigma_{\pi N}(=\!m_{ud} g_S^s)$ against the choice of the fitting form.
To this end, we test the linear and quadratic extrapolations of Eq.~(\ref{eq:polynomial_extrapolation}) as well as those based on ChPT.
In $O(p^3)$ covariant ChPT~\cite{alarcon}, $\sigma_{\pi N}$ is given as 
\begin{equation}
\sigma_{\pi N} =
c_1 m_\pi^2
- \frac{3 g_A^2 m_\pi^3}{16 \pi^2 f_\pi^2 m_N} 
\Biggl[
\frac{3 m_N^2 - m_\pi^2}{\sqrt{4m_N^2 - m_\pi^2}} \arccos \frac{m_\pi}{2 m_N}
+ m_\pi \ln \frac{m_\pi }{m_N}
\Biggr].
\label{eq:chiral_perturbation_sigma_term}
\end{equation}
We use our result (\ref{eq:our_g_a}) for $g_A$.
The nucleon mass is fixed to the experimental value $m_N = 939$~MeV by ignoring higher order corrections.
The only fit parameter is $c_1$.

\begin{table}[b]
\begin{center}
\begin{tabular}{l|c|cccc}
\hline \hline
Fitting form & $\sigma_{\pi N}$ at $m_{\pi, {\rm phys}}$  [MeV] &$c_1$ [GeV$^{-1}$] & $c_2$ [GeV$^{-3}$] &$\chi^2$/d.o.f. \\
\hline
Linear fit of $\sigma_{\pi N}$ & 21(1)  & 1.1(1) & $-$ & 1.47 \\
Quadratic fit of $\sigma_{\pi N}$ & 26(3)  & 1.4(2) & -1.4(7) & 0.39 \\
\hline
Eq. (\ref{eq:chiral_perturbation_sigma_term})
 & 42(1)  & 3.0(1) & $-$ & 3.96 \\
Eq. (\ref{eq:chiral_perturbation_sigma_term_2})
 & 34(3) & 2.6(2) & 2.3(7) & 0.61 \\
\hline
Linear fit of $g_S^s$ & 23(2)  & $-$ & $-$ & $-$ \\
\hline
\end{tabular}
\end{center}
\caption{
Numerical results of chiral extrapolations of $\sigma_{\pi N}$.
}
\label{table:nucleon_scalar_charges_chiral_perturbation}
\end{table}

Figure~\ref{fig:extrapolation_sigma_term} shows the chiral extrapolations of $\sigma_{\pi N}$. Numerical results are summarized in Table~\ref{table:nucleon_scalar_charges_chiral_perturbation}, where we also put $\sigma_{\pi N}$ at $m_{\pi,{\rm phys}}$ estimated from the linear fit to $g_S^s$ and Eq.~(\ref{eq:sigma_term}).
We observe that the polynomial fits describe our data of $\sigma_{\pi N}$ with $\chi^2/{\rm d.o.f}\!\lesssim\!1.5$, though the coefficient of the quadratic term is consistent with zero.
The extrapolated values of $\sigma_{\pi N}$ are in good agreement with that from the linear extrapolation of $g_S^s$.

Similar to the extrapolation of $g_A$, the one-loop ChPT formula leads to a large value of $\chi^2/{\rm d.o.f.}\!\simeq\!4$.
As shown in Table~\ref{table:nucleon_scalar_charges_chiral_perturbation}, $\chi^2$ is largely reduced, and the extrapolated value of $\sigma_{\pi N}$ significantly changes by including an $O(p^4)$ analytic term into the fitting form 
\begin{equation}
\sigma_{\pi N} =
\mbox{``right-hand side of Eq.~(\ref{eq:chiral_perturbation_sigma_term})''}
+c_2 m_\pi^4.
\label{eq:chiral_perturbation_sigma_term_2}
\end{equation}
From these observations, we conclude that higher order corrections in the chiral expansion are not small in the simulation region of $m_\pi$, and employ the quadratic fit of $\sigma_{\pi N}$ to determine its value at $m_{\pi,{\rm phys}}$.
Its systematic uncertainty is estimated by comparing with the linear fit of $\sigma_{\pi N}$, which yields a statistically significant value of the linear coefficient $c_1$ and a reasonable value of $\chi^2/{\rm d.o.f}$.

\begin{figure}[b]
\centering
\includegraphics[width=0.60\textwidth,clip]{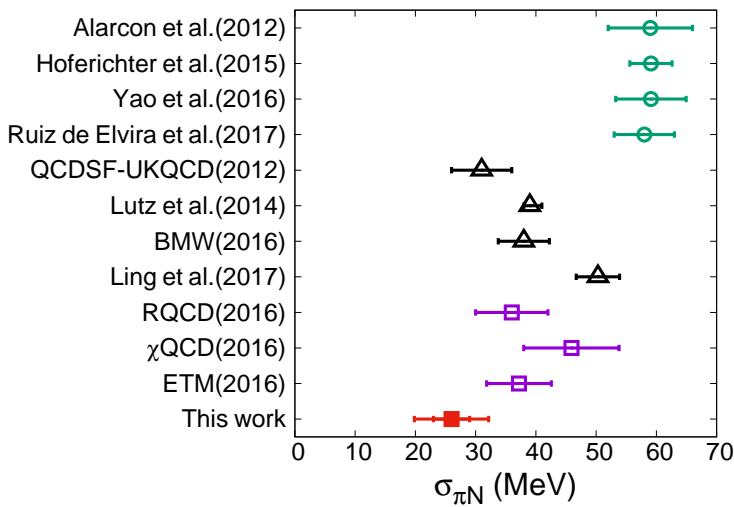}
\caption{\label{fig:sigma_term_lattice}
Our result for $\sigma_{\pi N}$ (filled square) compared with those from recent direct evaluations in lattice QCD (open squares, RQCD \cite{rqcdsigmaterm}, $\chi$QCD \cite{chiqcdsigmaterm}, ETM \cite{etmsigmaterm}), analyses of lattice QCD data using Feynman-Hellmann theorem (black triangles, QCDSF-UKQCD \cite{qcdsf-ukqcdsigmaterm}, Lutz {\it et al.} \cite{lutz}, BMW \cite{bmwsigmaterm}, Ling {\it et al.} \cite{ling}) and phenomenological studies (open circles, Alarc\'{o}n {\it et al.} \cite{alarcon}, Hoferichter {\it et al.} \cite{Hoferichter}, Yao {\it et al.} \cite{yao}, Ruiz de Elvira {\it et al.} \cite{deelvira}). 
As for our result, the smallest error bar denotes the statistical one, and the largest one also takes into account those due to the extrapolation and the discretization.
}
\end{figure}

\begin{figure}[tb]
\centering
\includegraphics[width=0.60\textwidth,clip]{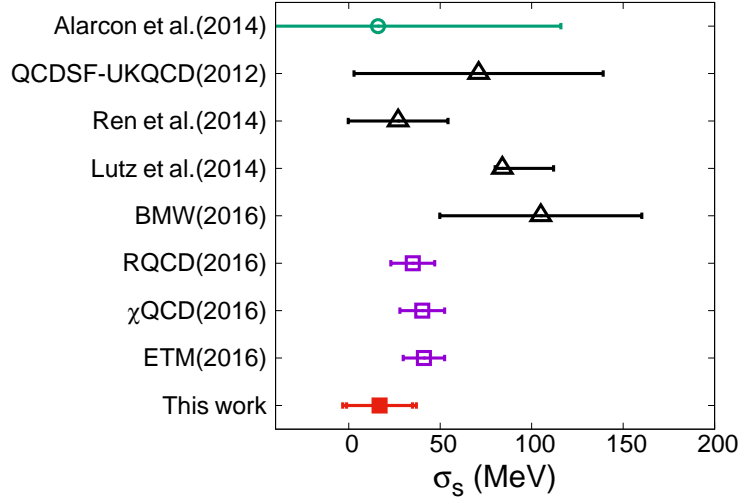}
\caption{\label{fig:strange_content_lattice}
Our result for $\sigma_s$ (filled square) compared with those from recent direct evaluations in lattice QCD (open squares, RQCD \cite{rqcdsigmaterm}, $\chi$QCD \cite{chiqcdsigmaterm}, ETM \cite{etmsigmaterm}), analyses of lattice QCD data using Feynman-Hellmann theorem (black triangles, QCDSF-UKQCD \cite{qcdsf-ukqcdsigmaterm}, Ren {\it et al.} \cite{ren}, Lutz {\it et al.} \cite{lutz}, BMW \cite{bmwsigmaterm}) and phenomenological studies (open circle, Alarc\'{o}n {\it et al.} \cite{alarcon}). 
The smallest error bar of our result denotes the statistical one, and the largest one also takes into account those due to the extrapolation and the discretization.
}
\end{figure}

Our numerical results are 
\begin{eqnarray}
g_S 
&=&
0.88 (8)_{\rm stat} (3)_{\chi} (7)_{a \neq 0}
,
\\
\sigma_{\pi N} 
&=&
26 (3)_{\rm stat} 
(5)_{\chi} 
(2)_{a \neq 0}~{\rm MeV}
,
\\
\sigma_s
&=&
17 (18)_{\rm stat} (9)_{\chi}~{\rm MeV}
,
\label{eq:strange_content_result}
\end{eqnarray}
where $O((a\Lambda_{\rm QCD})^2)$ discretization errors are assigned to $g_S$ and $\sigma_{\pi N}$. 
This error for $\sigma_s$ is much smaller than its total uncertainty and hence is neglected.
As shown in Fig.~\ref{fig:scalar_charge_extrapolation}, we observe good agreement with previous estimates of $g_S$~\cite{rqcdisovector,chiqcdisovector,pndmeisovector,etm2017,green,alonso} and $\sigma_s$~\cite{chiqcdsigmaterm,rqcdsigmaterm,bmwsigmaterm,etmsigmaterm,qcdsf-ukqcdsigmaterm,takeda,ohki2}.
Figure~\ref{fig:sigma_term_lattice} shows a comparison of $\sigma_{\pi N}$ with recent lattice~\cite{chiqcdsigmaterm,rqcdsigmaterm,bmwsigmaterm,etmsigmaterm,qcdsf-ukqcdsigmaterm} and phenomenological estimates~\cite{alarcon,Hoferichter,Hoferichter2,yao,ren,deelvira}.
We observe good agreement among lattice results, which are systematically smaller than the phenomenology as mentioned in the Introduction.
While we observe the slow convergence of Eq. (\ref{eq:chiral_perturbation_sigma_term}), recent phenomenological estimates slightly rely on ChPT by employing a dispersive analysis, and the necessary chiral correction is estimated to be small \cite{bernard2}.
Further studies are needed to resolve the tension with phenomenology.
We also show in Fig. \ref{fig:strange_content_lattice} the comparison of the results of the evaluations of the strange content of nucleon.
We see agreement between all results, although some are affected by large uncertainty.

It is worth noting that the disconnected diagram gives rise to a small contribution to $g_S^s$ and hence $\sigma_{\pi N}$.
Tables~\ref{table:nucleon_scalar_charges} and \ref{table:nucleon_scalar_charges_extrapolation} show that it is only 3\,--\,8\,\% contribution at simulated $m_\pi$'s, and this maintains down to $m_{\pi,{\rm phys}}$.
From the extrapolated value of $S_{u+d}^{\rm disc}$, the disconnected part $\sigma_{u+d}^{\rm disc}\!=\!m_{ud}S_{u+d}^{\rm disc}=2.0(2.3)$~MeV is only $8\pm9$\,\% contribution to $\sigma_{\pi N}$.
This is smaller than $O(1/N_c)$ expected from the large $N_c$ arguments and in favor of the OZI rule.

\section{\label{sec:tensor_analysis}Tensor charges}

For the tensor charges~(\ref{eq:tensor_charge}), we consider up, down and strange quark contributions, $\delta u$, $\delta d$ and $\delta s$, which are needed to study new physics effects to nucleon observables in the flavor basis.
We also report on the isovector tensor charge 
\begin{eqnarray}
g_T
&\equiv&
\frac{1}{2 m_N}
\langle p | \bar u i\sigma_{03} \gamma_5 u - \bar d i\sigma_{03} \gamma_5 d | p \rangle
= \delta u - \delta d
,
\end{eqnarray}
which has been studied in one-loop ChPT~\cite{detmold,chiralextrapolationtensor}
and previous lattice studies~\cite{rqcdisovector,pndmeisovector,chiqcdisovector,green,rbcukqcdisovectortensor,etm2017,nplqcd}.

\begin{figure}[tbp]
\centering
\includegraphics[width=0.48\textwidth,clip]{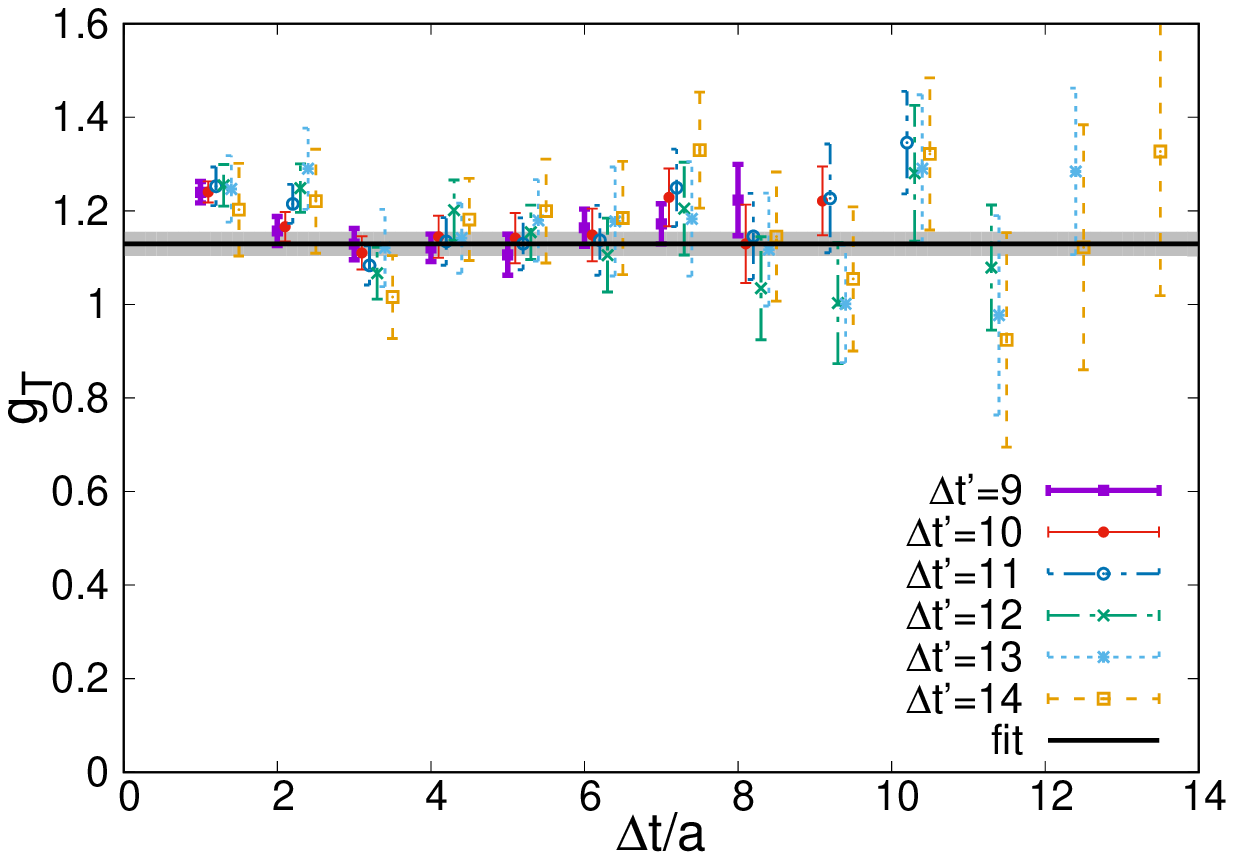}
\includegraphics[width=0.48\textwidth,clip]{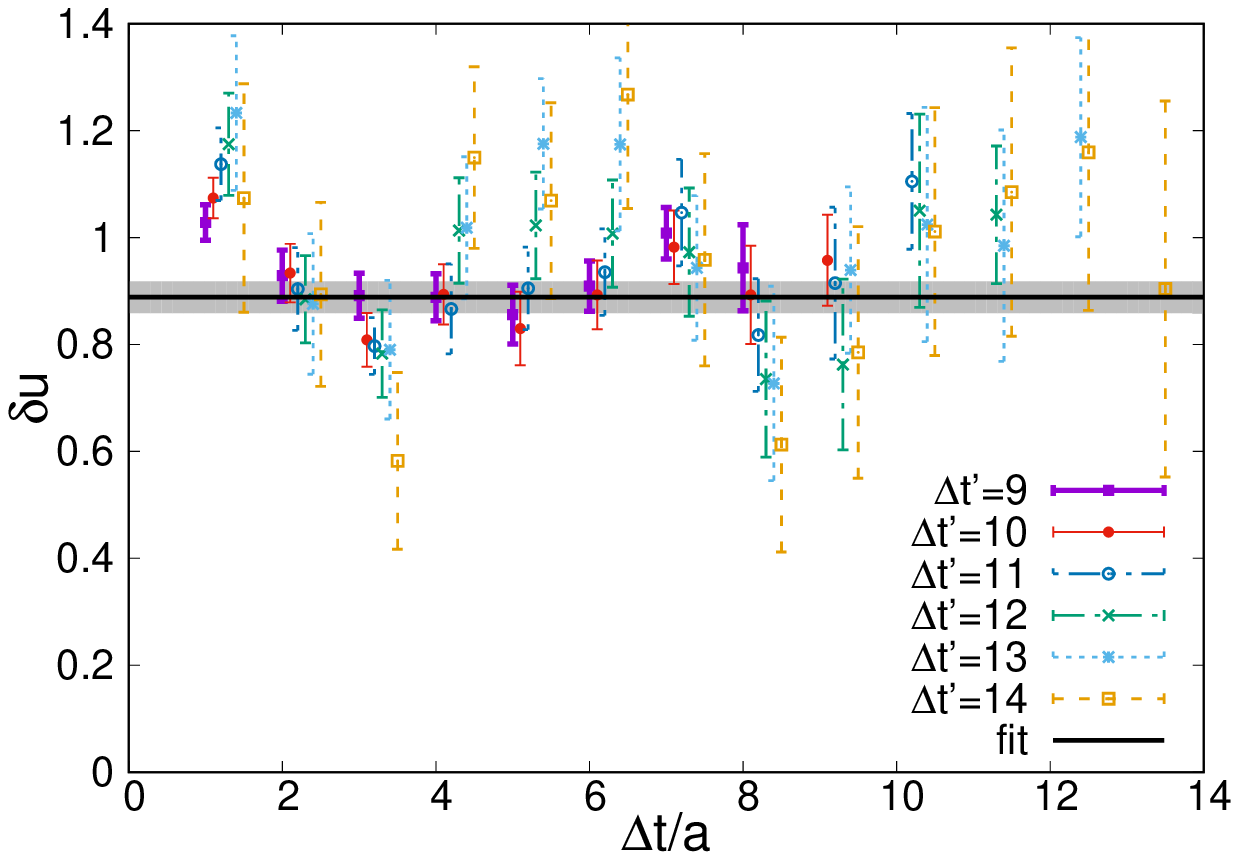}
\vspace{1mm}
\includegraphics[width=0.48\textwidth,clip]{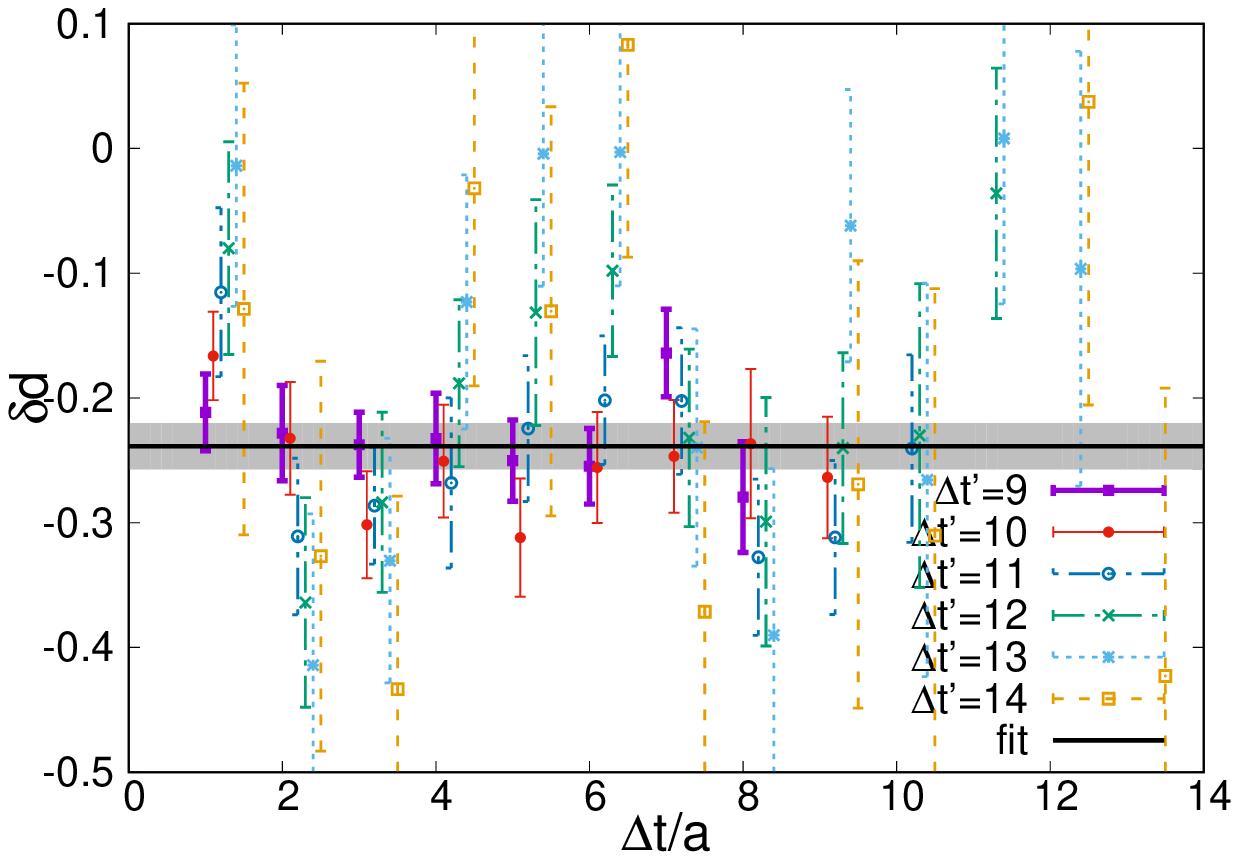}
\includegraphics[width=0.48\textwidth,clip]{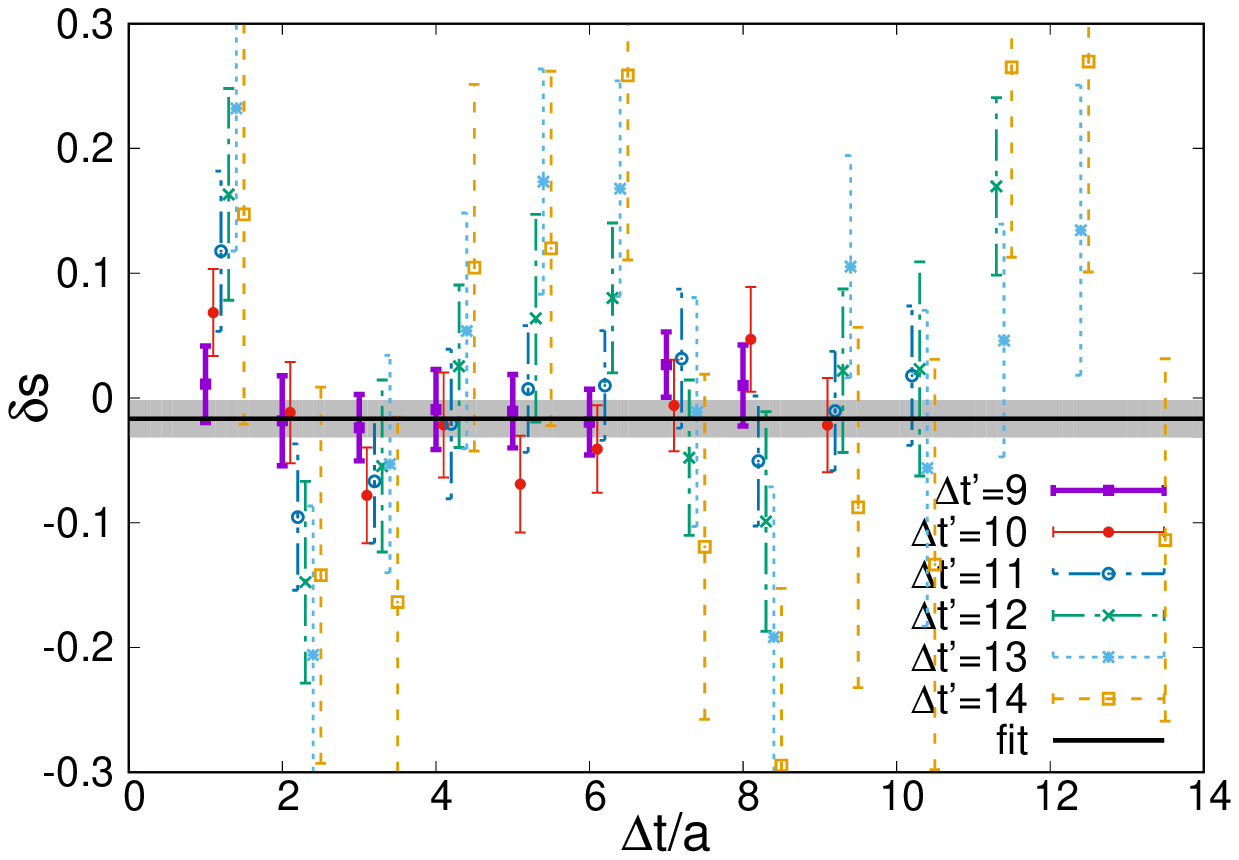}
\caption{\label{fig:tensor_charge_fit}
Effective values of tensor charges at $m_{ud}\!=\!0.015$ as a function of $\Delta t$.
The top-left panel shows our data of $g_T$, whereas top-right, bottom-left and bottom-right panels are for $\delta u$, $\delta d$ and $\delta s$, respectively.
}
\end{figure}

\begin{table}[tbp]
\begin{center}
\begin{tabular}{l|lccc}
\hline \hline
& $m_{ud}$ & $m_{\pi}$ (MeV) & Charge & $\chi^2$/d.o.f. \\
\hline
$g_T$ 
&0.050 & 540 & 1.215(29) & 0.29 \\
&0.035 & 450 & 1.195(31) & 0.76 \\
&0.025 & 380 & 1.123(18) & 1.27 \\
&0.015 & 290 & 1.129(26) & 0.53 \\
\hline
$\delta u $ 
&0.050 & 540 & 0.980(26) & 0.59 \\
&0.035 & 450 & 0.938(31) & 0.78 \\
&0.025 & 380 & 0.907(21) & 1.20 \\
&0.015 & 290 & 0.888(30) & 1.29 \\
\hline
$\delta d$ 
&0.050 & 540 & -0.233(16) & 0.65 \\
&0.035 & 450 & -0.261(19) & 0.63 \\
&0.025 & 380 & -0.222(17) & 1.02 \\
&0.015 & 290 & -0.239(19) & 1.29 \\
\hline
$\delta s$ 
&0.050 & 540 & -0.004(11) & 0.91 \\
&0.035 & 450 &  0.001(14) & 0.70 \\
&0.025 & 380 & -0.001(13) & 0.65 \\
&0.015 & 290 & -0.017(15) & 1.11 \\
\hline
\end{tabular}
\end{center}
\caption{
Numerical results for tensor charges from constant fit to $R_T (\Delta t,\Delta t^\prime)$ at simulation points.
}
\label{table:nucleon_tensor_charges}
\end{table}

Figure \ref{fig:tensor_charge_fit} shows the effective values of the tensor charges at $m_{ud}=0.015$. 
The Gaussian smearing works well to obtain plateaux, from which we determine the tensor charges by the constant fit in $\Delta t$ and $\Delta t^\prime$.
Numerical results are summarized in Table~\ref{table:nucleon_tensor_charges}.
$\chi^2/{\rm d.o.f.}\!<\!1.3$ at all simulation points.
The isovector charge $g_T$ is a purely connected contribution, and is determined with an accuracy of a few percent.
We observe that disconnected contributions to $\delta u$ and $\delta d$ are not large, similar to the case of $g_S^s$.
Their statistical accuracy is typically 3\% and 10\%, respectively.
On the other hand, $\delta s$ is a purely disconnected one, and is consistent with zero.

\begin{figure}[tbp]
\centering
\includegraphics[width=0.48\textwidth,clip]{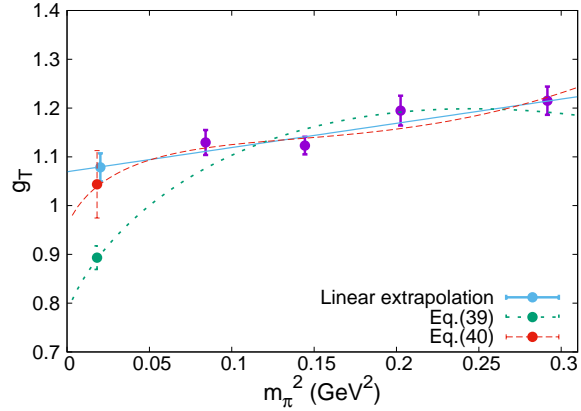}
\caption{\label{fig:extrapolation_isovector_tensor_charge}
Chiral extrapolation of $g_T$ using ChPT-based fitting forms of Eqs.~(\ref{eq:chiral_perturbation_gt}) (dotted line) and (\ref{eq:chiral_perturbation_gt_improved}) (dashed line).
The solid line shows the linear fit, from which we obtain the result in Eq.~(\ref{eq:tensor:g_T}).
}
\end{figure}

\begin{table}[htb]
\begin{center}
\begin{tabular}{l|l|c|cccc}
\hline \hline
& Fitting form & Charge at $m_\pi = 135$ MeV &$c_0$ &$c_1$&$c_2$& $\chi^2$/d.o.f. \\
\hline
$g_T $ 
& Constant & 1.15(1) & 1.15(1) & $-$ & $-$ & 3.35 \\
& Linear & 1.08(3) & 1.07(3) & 0.50(18) & $-$ & 1.09\\
& Quadratic & 1.11(7) & 1.12(9) & -0.1(1.0) & 1.5(2.6) & 1.85\\
& Eq.~(\ref{eq:chiral_perturbation_gt}) & 0.89(2) & 0.78(2) & -0.62(21) & $-$ & 4.18 \\
& Eq.~(\ref{eq:chiral_perturbation_gt_improved}) & 1.04(7) & 0.95(8) & -3.8(1.3) & 7.6(3.1) & 2.37 \\
\hline
$\delta u $
& Constant & 0.926(13) & 0.926(13) &$-$&$-$ & 2.26\\
& Linear & 0.853(31) & 0.844(34) & 0.46(18) &$-$& 0.04 \\
& Quadratic & 0.868(79) & 0.864(98) & 0.23(1.12) & 0.6(2.8) & 0.03 \\
\hline
$\delta d $ 
& Constant & -0.237(9) & -0.237(9) & $-$ & $-$ & 0.83 \\
& Linear & -0.235(21) & -0.235(23) & -0.01(11) & $-$ & 1.23 \\
& Quadratic & -0.215(49) & -0.209(61) & -0.33(71) & 0.8(1.8) & 2.27 \\
\hline
$\delta s $ 
& Constant & -0.004(6) & -0.004(6) &$-$&$-$ & 0.30 \\
& Linear & -0.012(16) & -0.012(17) & 0.04(8) &$-$& 0.33 \\
& Quadratic & -0.040(39) & -0.048(48) & 0.47(55) & -1.1(1.4) & 0.02 \\
\hline
\end{tabular}
\end{center}
\caption{
Numerical results of polynomial chiral extrapolations of tensor charges.
For $g_T$, we also list results with ChPT-based fitting forms~(\ref{eq:chiral_perturbation_gt}) and (\ref{eq:chiral_perturbation_gt_improved}).
}
\label{table:nucleon_tensor_charges_extrapolation}
\end{table}

The one-loop ChPT formula for $g_T$ is available in Refs.~\cite{detmold,chiralextrapolationtensor}. 
We fit our data to the formula of Ref. \cite{chiralextrapolationtensor}
\begin{equation}
g_T
=
c_0 \Biggl\{
1+\frac{m_\pi^2}{4 (4\pi f_\pi )^2} \Bigl[
(1+8 g_A^2) \ln \Bigl( \frac{\mu^2}{m_\pi^2 } \Bigr)
+2 +3g_A^2
\Bigr]
\Biggr\}
+ c_1 m_\pi^2
,
\label{eq:chiral_perturbation_gt}
\end{equation}
where $c_0$ and $c_1$ are fit parameters.
We set the renormalization scale to $\mu\!=\!770$~MeV, and $g_A$ to our result~(\ref{eq:our_g_a}).
This extrapolation is shown in Fig.~\ref{fig:extrapolation_isovector_tensor_charge}.
As seen in the numerical results in Table \ref{table:nucleon_tensor_charges_extrapolation}, this fit poorly describes our data with $\chi^2/{\rm d.o.f.}\!\gtrsim\!4$.
We therefore include an $O(p^4)$ analytic term into the fitting form
\begin{equation}
g_T
=
\mbox{``right-hand side of Eq.~(\ref{eq:chiral_perturbation_gt})''}
+ c_2 m_\pi^4 
.
\label{eq:chiral_perturbation_gt_improved}
\end{equation}
This fit is also shown in Fig.~\ref{fig:extrapolation_isovector_tensor_charge}.
The value of $\chi^2/{\rm d.o.f.}$ is significantly reduced to 2.4, but is still rather large.

\begin{figure}[tbp]
\centering
\includegraphics[width=0.48\textwidth,clip]{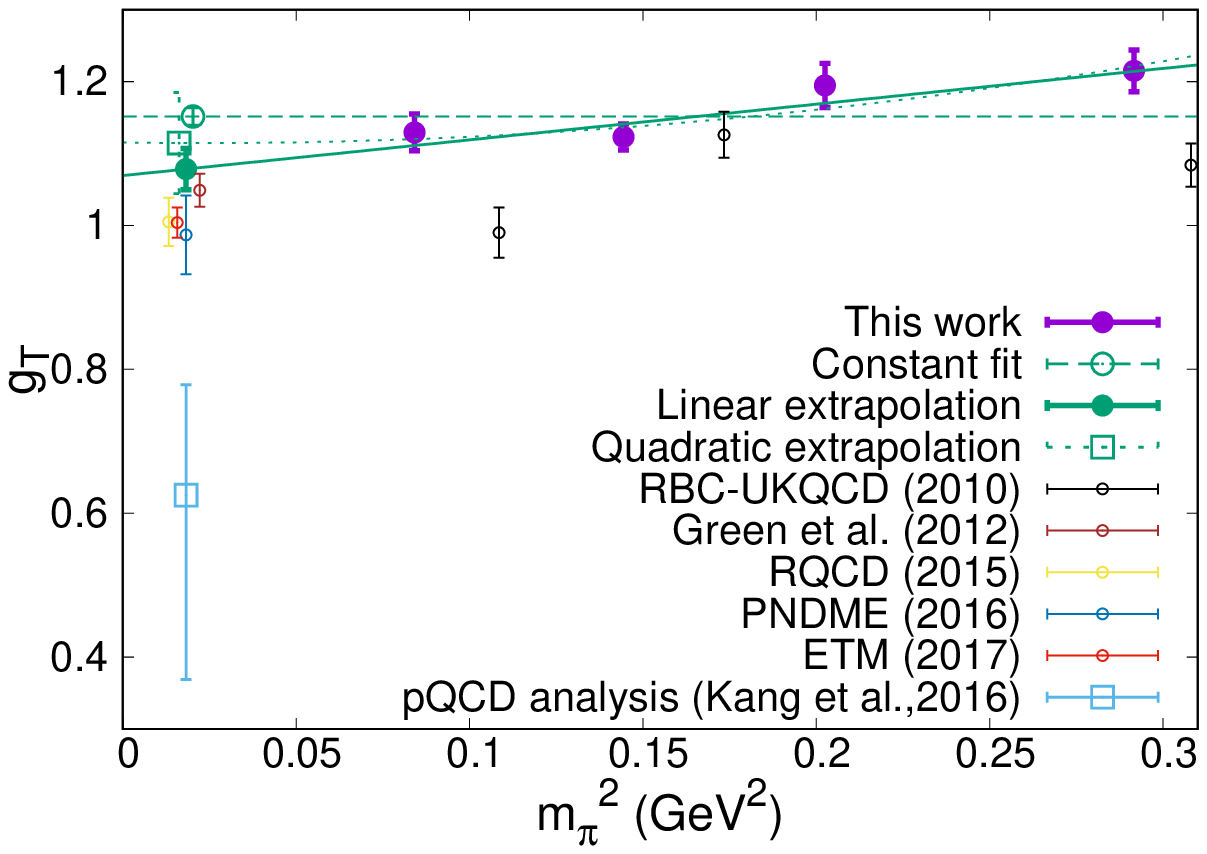}
\includegraphics[width=0.48\textwidth,clip]{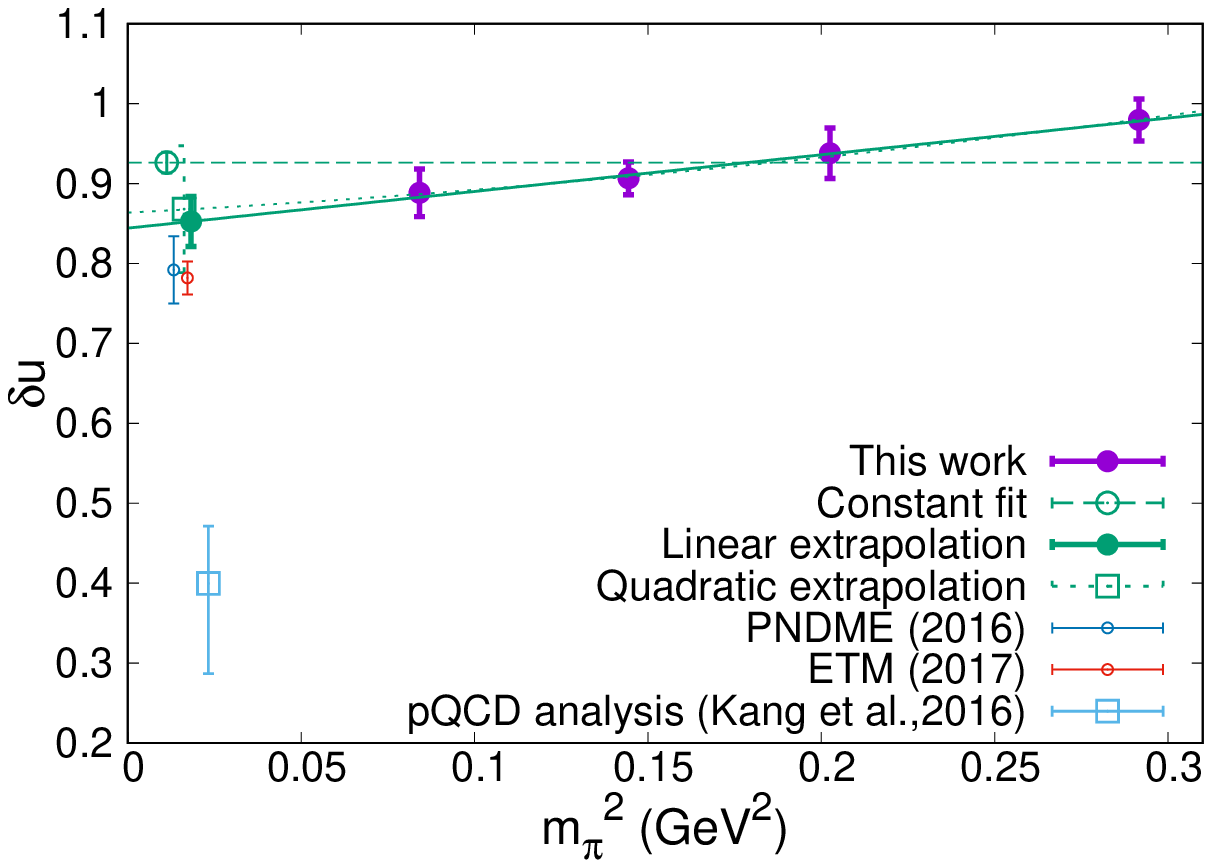}
\vspace{1mm}
\includegraphics[width=0.48\textwidth,clip]{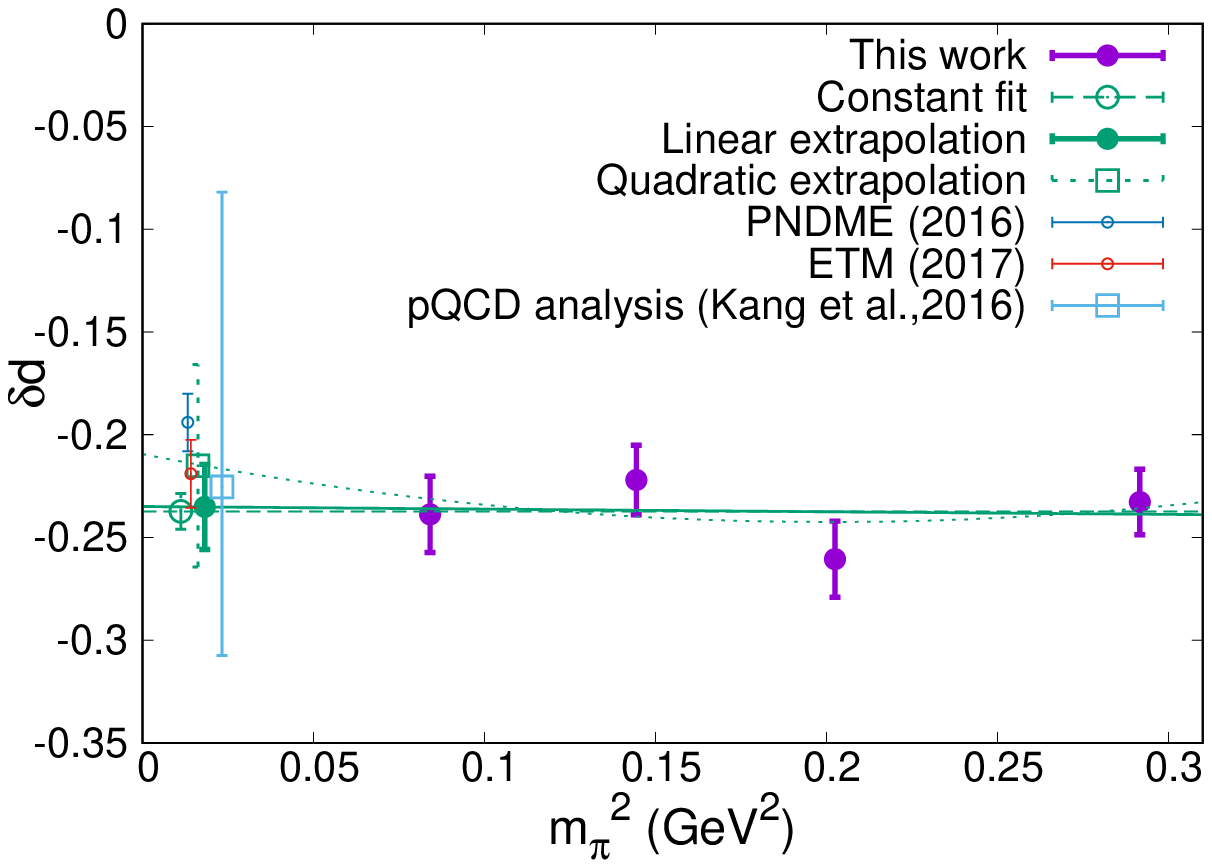}
\includegraphics[width=0.48\textwidth,clip]{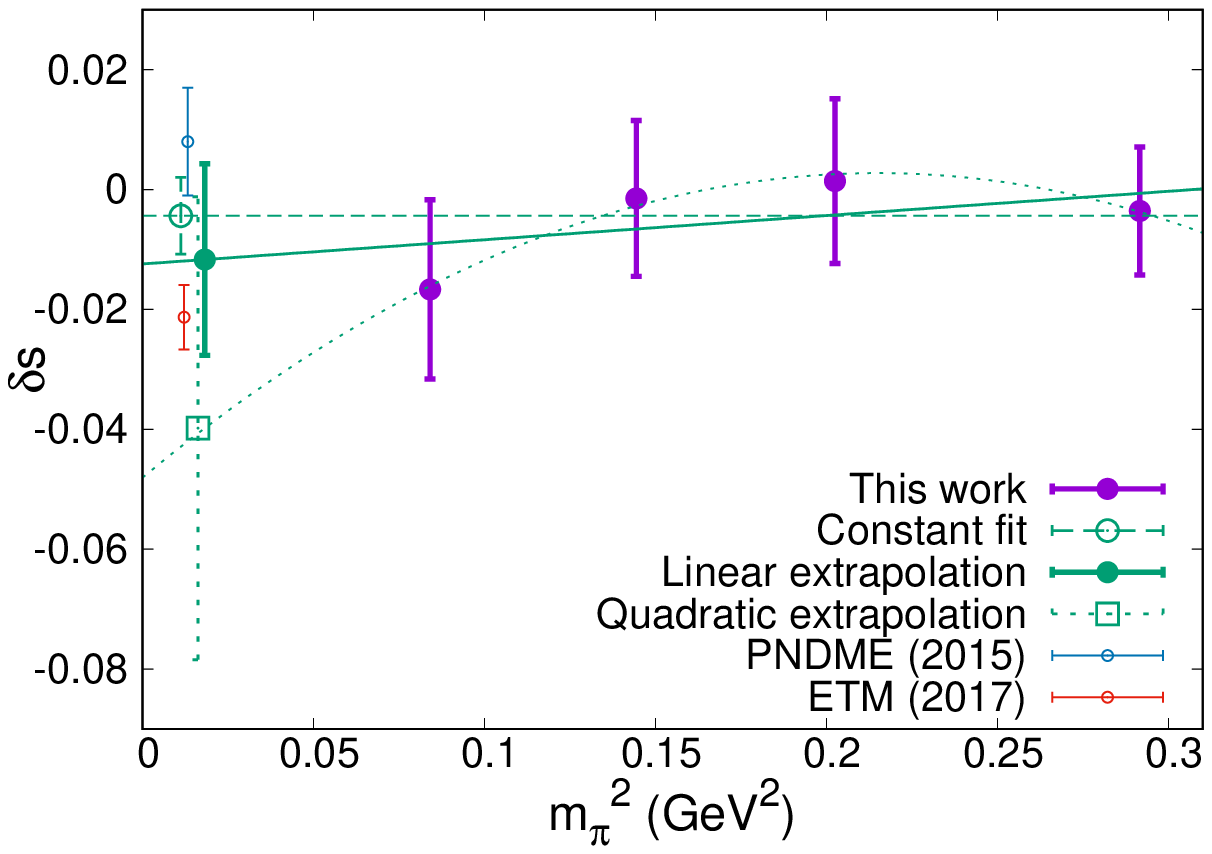}
\caption{\label{fig:tensor_charge_extrapolation}
Polynomial chiral extrapolations of tensor charges.
The top-left panel shows the extrapolation of $g_T$, whereas the top-right, bottom-left and bottom-right panels are for $\delta u$, $\delta d$ and $\delta s$, respectively.
Our data are plotted by filled circles, and the dashed, solid, and dotted lines show the constant, linear and quadratic extrapolations, respectively.
Small open circles are recent lattice estimates for $g_T$ (RQCD \cite{rqcdisovector}, PNDME \cite{pndmeisovector}, ETM \cite{etm2017}, Green {\it et al.} \cite{green}, RBC-UKQCD \cite{rbcukqcdisovectortensor}), $\delta u$, $\delta d$~\cite{pndmetensor1,pndmetensor2,etm2017}, and $\delta s$~\cite{pndmetensor1,pndmetensor2,etm2017}.
We also plot $g_T$, $\delta u$ and $\delta d$ from a perturbative QCD (pQCD) analysis of experimental data~\cite{kang} by blue open squares.
We note that we changed the renormalization scale of the data of Ref. \cite{kang} from $\mu^2 =10$ GeV$^2$ to $\mu = 2$ GeV according to the one-loop level renormalization \cite{tensorrenormalization}.
}
\end{figure}

We therefore extrapolate the tensor charges to $m_{\pi,{\rm phys}}$ using the polynomial parametrization~(\ref{eq:polynomial_extrapolation}).
Numerical results are summarized in Table~\ref{table:nucleon_tensor_charges_extrapolation}.
Chiral extrapolations are plotted in Fig.~\ref{fig:tensor_charge_extrapolation}.
Here we also show the results of the extractions from perturbative analysis \cite{kang}
\begin{eqnarray}
\delta u 
&=&
0.39^{+0.07}_{-0.11}
, 
\label{eq:tensorukang}
\\
\delta d 
&=&
-0.22^{+0.14}_{-0.08}
,
\label{eq:tensordkang}
\end{eqnarray}
at the renormalization scale $\mu^2 = 10$ GeV$^2$. 
The above tensor charges are consistent with a more recent extraction in the framework of collinear factorization \cite{radici2}.
Similar to the case of the axial and scalar charges, all the tensor charges show mild $m_\pi^2$ dependence, and hence the linear fit leads to reasonable values of $\chi^2/{\rm d.o.f.}\!\lesssim\!1$.
The coefficient $c_2$ of the quadratic term is poorly determined for all the charges, whereas even the constant fit works well for $\delta d$ and $\delta s$.
We therefore employ the linear fit to extrapolate the tensor charges.
The systematic error due to this choice of the fitting form is estimated by comparing with the quadratic fit for $g_T$ and $\delta u$ (same manner as $\sigma_{\pi N}$ with nonzero $c_1$) or the constant fit for $\delta d$ and $\delta s$ (same as other charges with $c_1\!\approx\!0$).

Numerical results for the tensor charges are 
\begin{eqnarray}
g_T
&=&
1.08 (3)_{\rm stat} (3)_{\chi} (9)_{a\neq 0}
\label{eq:tensor:g_T}
,
\\
\delta u
&=&
0.85 (3)_{\rm stat} (2)_{\chi} (7)_{a\neq 0}
,
\\
\delta d
&=&
-0.24 (2)_{\rm stat} (0)_{\chi} (2)_{a\neq 0}
,
\\
\delta s
&=&
-0.012 (16)_{\rm stat} (8)_{\chi} 
,
\end{eqnarray}
where $O((a\Lambda_{\rm QCD})^2)$ discretization errors for $\delta s$ are much smaller than the total uncertainty, and are neglected.
The isovector charge $g_T$ has been studied by previous lattice simulations, whereas only a few results are available for $\delta u$, $\delta d$, and $\delta s$ at the physical point~\cite{pndmetensor1,pndmetensor2,etm2017}.
We note that an analysis of experimental data of the quark transversity based on perturbative QCD led to significantly smaller $\delta u$ and, hence, $g_T$~\cite{kang}.
The analysis however suffers from an uncertainty due to the parametrization of the transversity at a momentum fraction $x$ inaccessible to the current experiments. 
Future experiments \cite{yez} will explore much wider region of $x$ and may resolve this discrepancy.

The strange quark contributions to the axial, scalar and tensor charges, namely $\Delta s$, $S_s$, and $\delta s$, are consistent with zero within our accuracy. 
The upper bound on $\delta s$ is, however, smaller than the other two.
This may be related to the fact that, in perturbative QCD, the disconnected contribution to the tensor charge requires at least three gluons connecting the quark loop and valence quarks due to Furry's theorem, whereas two gluons are enough for the axial and scalar charges.

\section{\label{sec:conclusion}Conclusion}

In this paper, we present our calculation of the nucleon scalar, axial, and tensor charges in $N_f=2+1$ flavor QCD.
We separately estimate the up, down and strange quark contributions to the charges by calculating the relevant disconnected diagrams using the all-to-all quark propagator.
Chiral symmetry is exactly preserved by employing the overlap quark action to suppress unphysical mixing among different flavors.
This simplifies the renormalization of the scalar and tensor charges, and allows us to avoid the contamination to the small strange quark contributions from the light quark ones.
We also employ the LMA and TSM to improve the statistical accuracy.

At the simulation points, the isovector charges $g_A$ and $g_T$ are determined with the statistical accuracy of a few percent, while it increases to 10\% for $g_S$ due to larger statistical fluctuation and cancellation between the up and down quark contributions.
The isoscalar charges and charges in the flavor basis receive disconnected contributions,
which give rise to a large uncertainty in $\Delta \Sigma$.
However, the disconcerted contributions turn out to be not large for the scalar ($g_S^s$ and hence $\sigma_{\pi N}$) and tensor ($\delta u$ and $\delta d$) charges, which are determined with an accuracy of $\approx$~10\,\%.
On the other hand, the strange quark contributions, $\Delta s$, $S_s$, and $\delta s$, are purely disconnected ones, and are consistent with zero within the statistical accuracy.
Except $\sigma_{\pi N}$ which explicitly contains $m_{ud}$ in its definition, we observe mild $m_{\pi}^2$ dependence of the nucleon charges in our region of $m_\pi\!\sim\!300$\,--\,500~MeV.
One-loop ChPT formulae poorly describe our data of $g_A$, $\sigma_{\pi N}$ and $g_T$, since the one-loop chiral logarithm leads to a too strong curvature to describe the mild $m_\pi^2$ dependence.
We therefore employ simple polynomial extrapolations, and observe reasonable consistency of previous lattice estimates with our results at the physical point.
We also observe reasonable consistency with experimental ($\Delta \Sigma$ and $\Delta s$) and phenomenological ($g_S$) estimates, whereas a perturbative QCD estimate of $\delta u$ and $g_T$ is significantly deviated from our and other lattice estimates.
The cause of this discrepancy is to be understood.

Our result for $g_A$ is marginally consistent with the experimental value within $O((a\Lambda_{\rm QCD})^2))$ $\approx\!8$\,\% discretization errors.
The ChPT formulae are expected to become a better guiding principle
of chiral extrapolation toward the chiral limit.
For a more precise and reliable determination of the isovector charges, it is interesting to extend this study to finer lattices and to lighter pion masses.
Our simulations in this direction are ongoing~\cite{JLQCD:Noaki:Lat14} by using a computationally inexpensive fermion formulation with good chiral symmetry~\cite{kanekomobius}.
On the other hand, the accuracy of $\Delta \Sigma$ and the strange quark contributions is limited by the statistical uncertainty. 
An improved calculation of the disconnected contributions is challenging but important 
toward the continuum limit and the physical point, but a better understanding of the proton spin puzzle and the search for new physics.

\begin{acknowledgments}
NY thanks Jordy de Vries, Takumi Doi, Nicolas Garron, Takumi Iritani, and V\'{e}ronique Bernard for useful discussion and comments.
The numerical calculations were performed on Hitachi SR16000 at High Energy Accelerator Research Organization under a support of its Large Scale Simulation Program (No.~16/17-14), and on Hitachi SR16000 at Yukawa Institute of Theoretical Physics.
This work is in part supported by the Grant for Scientific Research 
[Priority Areas ``New Hadrons'' (E01:21105006), (C) No.23540306] 
from the Ministry of Education, Culture, Science and Technology (MEXT) of Japan,
and as ``Priority Issue on Post-K computer''
(Elucidation of the Fundamental Laws and Evolution of the Universe) 
by MEXT and Joint Institute for Computational Fundamental Science (JICFuS).
This is also supported in part by JSPS KAKENHI Grant Numbers 
JP26247043, JP26400259, 17K14309, 
RIKEN iTHES Project, RSF grant  15-12-20008, 
RIKEN Special Postdoctoral Researcher program, 
Nara Women's University Intramural Grant for Project Research, 
and JSPS Postdoctoral Fellowships for Research Abroad.
\end{acknowledgments}

\end{document}